\documentclass[reprint,superscriptaddress,preprintnumbers,longbibliography, 
amsmath,amssymb,aps,superscriptaddress,floatfix,tightenlines,pra,
]{revtex4-2}

\usepackage{subfigure}
\usepackage{mathrsfs}
\usepackage[T1]{fontenc}
\usepackage{graphicx}
\usepackage{dcolumn}
\usepackage{bm}
\usepackage{color}
\usepackage{times}
\usepackage[colorlinks=true,
citecolor=magenta,
linkcolor=magenta,
anchorcolor=black,
urlcolor=magenta]{hyperref}
\usepackage{siunitx}
\usepackage{overpic}
\usepackage{braket}
\usepackage{changes}
\definechangesauthor[name={syh},color=blue]{syh}

\usepackage{enumerate}

\usepackage{amsthm}
\usepackage{amsmath}
\usepackage{amssymb}
\usepackage{amsfonts}
\theoremstyle{plain}

\newtheorem*{definition*}{Definition}

\newtheorem*{theorem*}{Theorem}
\newtheorem*{corollary*}{Corollary}
\newtheorem*{lemma*}{Lemma}
\newtheorem*{proposition*}{Proposition}

\NewDocumentEnvironment{variant}{O{theorem} D(){} m}
    {\addtocounter{#1}{-1}%
    \expandafter\renewcommand\csname the#1\endcsname{\ref{#3}$'$}%
    \begin{#1}[#2]}
    {\end{#1}}

\NewDocumentEnvironment{appthm}{O{theorem} D(){} m}
    {\addtocounter{#1}{-1}%
    \expandafter\renewcommand\csname the#1\endcsname{\ref{#3}}%
    \begin{#1}[#2]}
    {\end{#1}}

\begin{document}
    \title{Noisy Probabilistic Error Cancellation and Generalized Physical Implementability}

    \author{Tian-Ren Jin}
    \affiliation{Institute of Physics, Chinese Academy of Sciences, Beijing 100190, China}
    \affiliation{School of Physical Sciences, University of Chinese Academy of Sciences, Beijing 100049, China}

    \author{Yu-Ran Zhang}
    \email{yuranzhang@scut.edu.cn}
    \affiliation{School of Physics and Optoelectronics, South China University of Technology, Guangzhou 510640, China}

    \author{Kai Xu}
    \affiliation{Institute of Physics, Chinese Academy of Sciences, Beijing 100190, China}
    \affiliation{School of Physical Sciences, University of Chinese Academy of Sciences, Beijing 100049, China}
    \affiliation{Beijing Academy of Quantum Information Sciences, Beijing 100193, China}
    \affiliation{Hefei National Laboratory, Hefei 230088, China}
    \affiliation{Songshan Lake Materials Laboratory, Dongguan 523808, China}
    \affiliation{CAS Center for Excellence in Topological Quantum Computation, UCAS, Beijing 100190, China}

    \author{Heng Fan}
    \email{hfan@iphy.ac.cn}
    \affiliation{Institute of Physics, Chinese Academy of Sciences, Beijing 100190, China}
    \affiliation{School of Physical Sciences, University of Chinese Academy of Sciences, Beijing 100049, China}
    \affiliation{Beijing Academy of Quantum Information Sciences, Beijing 100193, China}
    \affiliation{Hefei National Laboratory, Hefei 230088, China}
    \affiliation{Songshan Lake Materials Laboratory, Dongguan 523808, China}
    \affiliation{CAS Center for Excellence in Topological Quantum Computation, UCAS, Beijing 100190, China}

    \begin{abstract}
    Decoherence severely limits the performance of quantum processors, posing challenges to reliable quantum computation. Probabilistic error cancellation, a quantum error mitigation method, counteracts noise by quasiprobabilistically simulating (non-physical) inverse noise operations. However, existing formulations of physical implementability, quantifying the minimal cost of simulating non-physical operations using physical channels, do not fully account for the experimental constraints, since noise also affects the cancellation process and not all physical channels are experimentally accessible. Here, we generalize the physical implementability to encompass arbitrary convex sets of experimentally available quantum states and operations. Within this generalized framework, we demonstrate noiseless error cancellation with noisy Pauli operations and analyze the bias of noisy cancellation. Furthermore, we establish connections between generalized physical implementability and quantum information measures, e.g., diamond norm, logarithmic negativity, and purity. These findings enhance the practical applicability of probabilistic error cancellation and open new avenues for robust quantum information processing and quantum computing.
    \end{abstract}
    \maketitle

    \noindent
    \textbf{\Large{Introduction}}

    \noindent
    In quantum computation, an ideal quantum circuit is unitary~\cite{nielsen2010quantum}.
    However, the imperfection of quantum devices will lead to the noises in the performance of quantum circuits.
    The physical operations on a quantum system are thought to be completely positive and trace-preserving (CPTP), which is also called the quantum channel~\cite{RevModPhys.91.025001}.
    The Markov noise in quantum circuit can be depicted as a quantum noise channel $\mathcal{E}$~\cite{breuer2002theory}.

    In practice, the noise channel $\mathcal{E}$ can be evaluated by the quantum process tomography, even though the cost may be exponentially overwhelming~\cite{PhysRevA.77.032322}.
    With the knowledge of the noise channel $\mathcal{E}$, we would like to implement its inverse $\mathcal{E}^{-1}$ to cancel the impact of the noise.
    However, only the unitary channels have quantum channels as its inverse $\mathcal{E}^{-1}$ is also $\textrm{CPTP}$~\cite{wigner2012group}.
    It is impossible to physically implement a quantum channel to cancel the incoherence error.

    Although the inverse operation $\mathcal{E}^{-1}$ may not be a quantum channel but rather a Hermitian-preserving and trace-preserving (HPTP) operation, it can be simulated with a series of quantum channels.
    If it can be decomposed as the affine combination of quantum channels, the inverse operation $\mathcal{E}^{-1}$ can be simulated by the quasiprobability mixture, of which the absolute value of coefficients in the affine decomposition are normalized into a probabilistic distribution.
    The noise inverse operation is the probabilistic mixture of quantum channels with signatures of coefficients in this distribution up to normalization.
    This quantum error mitigation method is the probabilistic error cancellation (PEC)~\cite{RevModPhys.95.045005,PhysRevLett.119.180509,PhysRevX.8.031027,cai2021multi,PhysRevResearch.3.033178,takagi2022fundamental,PhysRevApplied.15.034026,PRXQuantum.3.010345,PRXQuantum.2.040330,PRXQuantum.3.040313,piveteau2022quasiprobability,van2023probabilistic}. 

    Furthermore, the reduced dynamics with correlated initial conditions might not always be CPTP~\cite{PhysRevLett.73.1060,PhysRevA.77.042113}.
    The quasiprobability mixture technique allows for simulating such HPTP operations with quantum channels, which helps to analyze the non-Markov noise.
    The logarithm of the minimal cost, or overhead $C_{em}$ in the context of quantum error mitigation~\cite{PhysRevResearch.3.033178,cai2021practical,RevModPhys.95.045005}, to simulate an HPTP operation with CPTP channels is defined as the physical implementability of the HPTP operation~\cite{PhysRevResearch.3.033178,Jiang2021physical}.
    It has also been shown that the physical implementability of the noise inverse operation $\mathcal{E}^{-1}$ characterizes the decoherence effects of noise channel~\cite{guo2023noise}.

    In the PEC method, by employing the Pauli-twirling technique~\cite{PhysRevA.94.052325,PhysRevX.7.021050,cai2019constructing,PhysRevX.11.041039,PhysRevX.11.031057,van2023probabilistic,RevModPhys.95.045005}, the error channel and the noises in experiments can be randomized compiled to be Pauli diagonal. 
    By employing the cycle benchmarking~\cite{erhard2019characterizing,PRXQuantum.3.020357} and error reconstruction technique~\cite{10.1145/3408039,harper2020efficient}, the error model of the noisy circuit can be constructed~\cite{PhysRevX.11.041039,van2023probabilistic,Ferracin2024efficiently}.
    In principle, with the well-performed randomized compiling and sufficient estimation of error model, the PEC method is believed to be free of bias~\cite{RevModPhys.95.045005}.
    However, in practice, the PEC method will not always be unbiased.
    The bias can be induced from the violation of the error model~\cite{PRXQuantum.4.040329,govia2024bounding}.
    This may be caused by the inaccurately randomized compiling or the inaccurate error model estimation~\cite{govia2024bounding}.

    Moreover, to cancel the error in noisy circuit,  additional Pauli gates need to be introduced in the implementation of the PEC, which would also be affected by the noises in the additional circuit.
    Intuitively, the noises on Pauli gates $\mathcal{P}_i$ may be single-qubit errors, which is not comparable to the error to be canceled.
    However, due to the cross-talk between qubits~\cite{PRXQuantum.3.020301}, the noises of multi-qubit Pauli gates may not be the tensor product of noises of single-qubit Pauli gates in experiments.
    Therefore, the noises introduced in Pauli gates are not simply the single-qubit errors, and it is possible to yield a considerable bias of cancellation.
    If the noises in the cancellation are less than the error to be canceled, the PEC has a positive effect on the performance of a quantum circuit.
    Otherwise, the cancellation makes the performance of the quantum circuit even worse.
    The similar imperfection in quantum error correction leads to the famous threshold theorem~\cite{knill1996threshold,10.1145/258533.258579,kitaev1997quantum,RevModPhys.87.307}, which may be the main obstacle to realizing fault-tolerant quantum computation.
    The error mitigation technique is not believed to have an inevitable threshold phenomenon~\cite{RevModPhys.95.045005}.
    Nevertheless, the noisy implementation of the cancellation will also lead to additional bias.

    In this paper, we investigate the noisy cancellation of the error channel.
    We generalized the physical implementability to an arbitrary given convex set $\mathcal{F}$ of quantum operations and states that are available in experiments, and its general properties are introduced in Methods.
    Using the generalized physical implementability, we demonstrate the optimal method to noiseless cancel a given error channel with noisy Pauli gates in Results.
    We also consider the bias of the noisy cancellation with the noisy Pauli gates and the bias of the error model violation.
    Moreover, we illustrate the connection between the generalized physical implementability and other quantum information measures, such as the diamond norm, logarithmic negativity, and purity in Discussion. 
    \\
    \\\textbf{\Large{Results}}\\
    \noindent \textbf{Noiseless Cancellation with Noisy Pauli Basis}
    \\
    Let $\mathcal{N}$ be an HPTP operation on system $A$, and then it can be decomposed as the affine combination of CPTP channels $\mathcal{N}_{i}$
    \begin{equation}
        \mathcal{N} = \sum_i n_i \mathcal{N}_{i},
    \end{equation}
    where $\sum_i n_i = 1$.
    If all $n_i \geq 0$, it can be implemented by a probabilistic mixture of CPTP channels $\mathcal{N}_{i}$.
    Otherwise, for any $n_i \leq 0$, we should rewrite it as 
    \begin{equation}
        \mathcal{N} = Z \sum_i \mathrm{sgn}(n_i) q_i \mathcal{N}_{i},
    \end{equation}
    where $Z = \sum_i |n_i|$ and $q_i = {|n_i|}/{Z}$.
    Then, $\mathcal{N}$ can be simulated by using the probabilistic mixture of CPTP channels $\mathcal{N}_{i}$ with sign $\mathrm{sgn}(n_i)$.
    The quantity $Z$ is the cost of CPTP channels for the implementation of a single HPTP operation.
    The physical implementability~\cite{Jiang2021physical} is defined as the logarithm of the minimal cost of CPTP operations for HPTP operations
    \begin{equation}
        \nu(\mathcal{N}) = \log \min \left\{\sum_i |n_i|: \mathcal{N} = \sum_i n_i \mathcal{N}_{i}, \mathcal{N}_{i} \in \mathcal{Q} \right\},
    \end{equation} 
    where $\mathcal{Q}$ denote the set of CPTP channels.

\begin{figure*}[t]
    \centering
    \includegraphics[width=\textwidth]{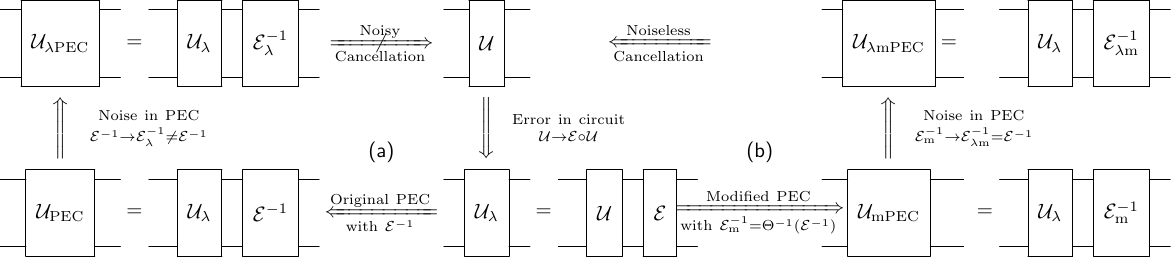}
    \caption{\textbf{Diagram of the PEC method and its modification.}
        \textbf{(a)}, The noisy implementation $\mathcal{U}_{\lambda}$ of a quantum circuit $\mathcal{U}$ with error channel $\mathcal{E}$.
        The PEC method cancels this error by implementing its inverse operation $\mathcal{E}^{-1}$.
        \textbf{(b)}, However, the noisy realization of the inverse operation $\mathcal{E}_{\lambda}^{-1}$ incompletely cancels the error $\mathcal{E}$.
        With the noises of cancellation, the PEC operation needs to be modified as $\mathcal{E}_{\mathrm{m}}^{-1} = \Theta^{-1}(\mathcal{E}^{-1})$, where $\Theta$ is the map from the ideal Pauli gates $\mathcal{P}_i$ to noisy gate $\mathcal{K}_i$, such that its noisy realization $\mathcal{E}_{\lambda\mathrm{m}}^{-1}$ cancels the error $\mathcal{E}$ complete.
    }
    \label{fig: diagram}
\end{figure*}

    For the PEC quantum error mitigation method, the error channel is mitigated by simulating its inverse operation. 
    Let $\mathcal{E} \equiv \mathcal{U}_{\lambda} \circ \mathcal{U}^{\dagger}$ be the error channel (in the left action) of an ideal quantum circuit $\mathcal{U}$ whose noisy circuit in experiments is $\mathcal{U}_{\lambda}$, as shown in Fig.~\ref{fig: diagram}.
    The quasiprobability decomposition of the inverse operation $\mathcal{E}^{-1}$ of error channel is 
    \begin{equation}
        \mathcal{E}^{-1} = \sum_i r_i \mathcal{P}_{i},
    \end{equation}
    where $\mathcal{P}_{i}$ are called noisy basis.
    With the Pauli-twirling technique, we consider the error model as the Pauli diagonal error, and the noisy basis $\mathcal{P}_{i}$ are Pauli gates.
    Then, the ideal expectation of the operator $\hat{O}$ relative to the initial state $\rho$ is 
    \begin{equation}
        \braket{\hat{O}}_0  = Z \sum_i \mathrm{sgn}(r_i) \frac{|r_i|}{Z} \mathrm{Tr}\left[\hat{O} \mathcal{P}_{i}\circ \mathcal{U}_{\lambda}(\rho)\right] .
    \end{equation}
    The PEC mitigated unitary circuit $\mathcal{U}_{\mathrm{PEC}}$ is shown in Fig.~\ref{fig: diagram}(a).

    Ideally, the Pauli gates $\mathcal{P}_{i}$ are physically realizable quantum channels, but the inevitable noises in experiments lead to the noisy Pauli gates $\mathcal{K}_i$.
    Here, we encounter two different kinds of noise, one is the error channel $\mathcal{E}$ in the quantum circuit $\mathcal{U}$ of our interest, and another is the noises in the noisy basis $\mathcal{P}_{i}$, which is used to cancel $\mathcal{E}$.
    To clearly distinguish them, the former one $\mathcal{E}$ is called error, while the latter is called noise in the following.
    The noisy realization of the inverse of the error channel thus is 
    \begin{equation}
        \mathcal{E}_{\lambda}^{-1} = \sum_i r_i \mathcal{K}_i \neq \mathcal{E}^{-1},
    \end{equation}
    which cannot cancel the error channel completely, as Fig.~\ref{fig: diagram}(a)

    Instead, if we ideally apply a modified PEC operation of the error channel, shown in Fig.~\ref{fig: diagram}(b), 
    \begin{equation}
        \mathcal{E}_{\mathrm{m}}^{-1} = \sum_i q_i \mathcal{P}_i,
    \end{equation}
    its noisy realization is 
    \begin{equation} \label{eq: inverse}
        \mathcal{E}_{\lambda\mathrm{m}}^{-1} = \sum_i q_i \mathcal{K}_i.
    \end{equation}
    We can select the parameters $q_i$ to cancel $\mathcal{E}$ completely, i.e. $\mathcal{E}_{\lambda\mathrm{m}}^{-1} = \mathcal{E}^{-1}$, based on the knowledge of the noises on Pauli gates.

    In precise, with the Pauli twirling techniques, we still assume that the noisy Pauli gates are Pauli diagonal
    \begin{equation}
        \mathcal{K}_i = \Theta(\mathcal{P}_i) = \sum_{j} \Theta_{ij} \mathcal{P}_j.
    \end{equation}
    The noise map $\Theta$ can be calibrated on the experimental devices. 
    If the noisy Pauli gate $\mathcal{K}_i$ are not completely indistinguishable, i.e. $\mathcal{K}_i$ are linearly independent, the noise linear map $\Theta$ is invertible.
    Then, the modified PEC operation of the error channel is 
    \begin{equation} \label{eq: pec_noisy}
        \mathcal{E}_{\mathrm{m}}^{-1} = \Theta^{-1}(\mathcal{E}^{-1}) = \sum_i q_i \mathcal{P}_i,
    \end{equation}
    where $q_i = \sum_j r_j \Theta^{-1}_{ji}$. 

    There still exists a problem that which model of the noisy quasiprobability cancellation of error $\mathcal{E}$ is optimal.
    To depict the cost of noisy cancellation, We thus define the generalized physical implementability for any HPTP operation $\mathcal{N}$ with respect to noisy CPTP operations as 
    \begin{equation}
        p_{\Theta(\mathcal{Q})}(\mathcal{N}) = \inf \left\{ \sum_i |x_i| : \mathcal{N} = \sum_i x_i \mathcal{N}_{i}, \mathcal{N}_{i} \in \Theta(\mathcal{Q}) \right\}
    \end{equation}
    where the minimization is the quasiprobability decompostion $\mathcal{N} = \sum_i x_i \mathcal{N}_{i}$ with respect to  noisy CPTP channels $\Theta(\mathcal{Q})$.
    Here, $\Theta$ denotes the noise map for the basis of operation space.
    Moreover, for other quantum information processing tasks involving the quasiprobability decomposition, their optimal costs can also be quantified by several types of implementability functions similar to the generalized physical implementability as defined.  
    For the general definition and properties of the implementability function, see Methods.

    With the generalized physical implementability, the optimal cost of the inverse operation $\mathcal{E}^{-1}$ is $p_{\Theta(\mathcal{Q})}(\mathcal{E}^{-1})$.
    Assume that the linear map $\Theta$ is invertible, by the affine invariance of the implementability function (see Methods), we have 
    \begin{equation}
        p_{\Theta(\mathcal{Q})}(\mathcal{E}^{-1}) = p_{\mathcal{Q}}(\Theta^{-1}(\mathcal{E}^{-1})) =  p_{\mathcal{Q}}(\mathcal{E}_{\mathrm{m}}^{-1}),
    \end{equation}
    where $p_{\mathcal{Q}} = \exp \nu$ is the (exponential) physical implementability.
    Since the optimal cancellation of a mixed-unitary channel with ideal CPTP channels is the decomposition with the unitaries~\cite{Jiang2021physical}, the optimal cancellation of modified PEC cancellation operation $\mathcal{E}_{\mathrm{m}}^{-1}$ is the quasiprobability decomposition with respect to the ideal Pauli channels $\mathcal{P}_i$ with quasiprobability
    \begin{equation} \label{eq: noise-free}
        q_i = \sum_j r_j \Theta^{-1}_{ji}.
    \end{equation} 
    The optimal cancellation of $\mathcal{E}^{-1}$ with respect to the noisy CPTP channels $\Theta(\mathcal{Q})$ thus is the quasiprobability decomposition with respect to the noisy Pauli channels $\mathcal{K}_i$ with $q_i$ in Eq.~(\ref{eq: noise-free}).
    In conclusion, with the matrix $\Theta_{ij}$ measured from the noisy Pauli basis $\mathcal{K}_i$, the inverse operation $\mathcal{E}^{-1}$ of the error channel can be optimally cancelled under the noisy Pauli basis $\mathcal{K}_i$.

    Here, we illustrate the result with simple examples.
    Assume the noise is a depolarizing error on one qubit, 
    ideally, the error channel of evolution with error rate $\lambda$ is 
    \begin{equation}
        \mathcal{E} = \left(1 - \frac{3\lambda}{4}\right)\mathcal{I} + \frac{\lambda}{4} (\mathcal{X} + \mathcal{Y} + \mathcal{Z}).
    \end{equation}
    It is not difficult to show that its inverse is
    \begin{equation} \label{eq: depolarizing}
        \mathcal{E}^{-1} = \frac{4 - \lambda}{4 (1 - \lambda)} \mathcal{I} - \frac{\lambda}{4 (1 - \lambda)} (\mathcal{X} + \mathcal{Y} + \mathcal{Z}).
    \end{equation} 
    If the error rate $\lambda \ll 1$, we approximate $\Theta(\mathcal{P}_i) \approx \mathcal{P}_i$, for $\mathcal{P}_i = \mathcal{X}, \mathcal{Y}, \mathcal{Z}$, we have
    \begin{equation} \label{eq: depolarizing_simu}
        \mathcal{U} = \frac{4 - \lambda}{4 (1 - \lambda)} \mathcal{U}_{\lambda} - \frac{\lambda}{4 (1 - \lambda)} (\mathcal{X} + \mathcal{Y} + \mathcal{Z}) \circ \mathcal{U}_{\lambda},
    \end{equation}
    which is in coincidence the known results of the depolarizing error~\cite{PhysRevLett.119.180509}.
    Then, we assume that $\Theta(\mathcal{I}) = \mathcal{I}$ and $\mathcal{K}_{\mathcal{P}_i} = \Theta(\mathcal{P}_i) = \mathcal{E}^{\alpha} \circ \mathcal{P}_i$ for $\mathcal{P}_i = \mathcal{X}, \mathcal{Y}, \mathcal{Z}$.
    It can be calculated as
    \begin{equation}
        \mathcal{E}^{\alpha} = \frac{1 + 3(1 - \lambda)^{\alpha}}{4} \mathcal{I} + \frac{1 - (1 - \lambda)^{\alpha}}{4} (\mathcal{X} + \mathcal{Y} + \mathcal{Z}).
    \end{equation} 
    Let $a = \frac{1 + 3(1 - \lambda)^{\alpha}}{4}, b = \frac{1 - (1 - \lambda)^{\alpha}}{4}$, we have 
    \begin{equation} \label{eq: noise_depolarizing}
        (\Theta_{ij}) = 
        \left(\begin{array}{cccc}
        1 & 0 & 0 & 0 \\
        b & a & b & b \\
        b & b & a & b \\ 
        b & b & b & a
        \end{array}\right).
    \end{equation}
    Thus, the quasiprobability, Eq.~(\ref{eq: noise-free}), of optimal cancellation in terms of $(\mathcal{K}_{I}, \mathcal{K}_{X}, \mathcal{K}_{Y}, \mathcal{K}_{Z})^{T}$ is obtained from Eqs.~(\ref{eq: depolarizing}) and~(\ref{eq: noise_depolarizing}).

    For an arbitrary Pauli diagonal error, since it is CPTP, in general, it can be written as the exponential of Lindblad operators~\cite{breuer2002theory}.
    \begin{equation}
        \mathcal{E} = \exp {\mathcal{L}},
    \end{equation}
    where the Lindblad operator can be written as
    \begin{equation}
        \mathcal{L} = \sum_{i} \lambda_i (\mathcal{P}_i-\mathcal{I}).
    \end{equation}
    Thus, the error channel is
    \begin{equation} \label{eq: Pauli_Lindblad}
        \mathcal{E} = \bigcirc_{i} [\omega_i \mathcal{I} + (1-\omega_i) \mathcal{P}_i],
    \end{equation}
    where $\omega_i = (1+\mathrm{e}^{-2\lambda_i})/2$.
    This error model is called the Pauli-Lindblad noise model~\cite{van2023probabilistic}.
    The inverse operation of the error channel is
    \begin{align}
        \mathcal{E}^{-1} & = \exp(-\mathcal{L}) = \bigcirc_{i} \left(\mu_i \mathcal{I} + (1-\mu_i) \mathcal{P}_i\right) \\
        & = \mathrm{e}^{2\sum_i\lambda_i} \bigcirc_{i} \left[\omega_i \mathcal{I} + (1-\omega_i) (-\mathcal{P}_i)\right] \nonumber
    \end{align}
    where $\mu_i = (1+\mathrm{e}^{2\lambda_i})/2$.
    There are two ways to simulate the inverse operation of the error channel.
    One is to simulate $\mathcal{E}^{-1}$ as a whole channel, and the other is to simulate each layer $\left(\mu_i \mathcal{I} + (1-\mu_i) \mathcal{P}_i\right)$, separately.
    No matter the ideal or noisy cancellations, the second way may have more cost than the first way~\cite{Jiang2021physical}, since the sub-multiplicity of the implementability function (see Methods).
    For the formalism simplicity, however, we only consider the noisy cancellation in the second way.
    By measuring the noise map $\Theta$ of Pauli gates in the experiment, the optimal noisy cancellation is given as
    \begin{equation}
        \mathcal{E}^{-1} = \bigcirc_{i} \left[\mu_i \mathcal{I} + (1-\mu_i) \sum_j \Theta^{-1}_{ij}\mathcal{K}_j\right].
    \end{equation}
    \\
    \\\noindent \textbf{Invertibility of Noise Map}
    \\
    The above discussion is based on the assumption that the linear map $\Theta$ is invertible.
    When the map $\Theta$ is not invertible, there is no perfect noisy cancellation.
    The invertibility of $\Theta$ is equivalent to $\det\Theta \neq 0$.
    In practice, assuming the true value $\Theta_0$ of the noise map is not invertible, i.e., $\det\Theta_0 = 0$, the random fluctuation from the finite measurements will lead to the noise map $\Theta$ measured from the experiments to be invertible, $\det\Theta \neq 0$.
    Therefore, the condition $\det\Theta \neq 0$ calculated with the experimental data does not certainly imply the invertibility of the noise map $\Theta$.
    In the following, we discuss the condition the noise map $\Theta$ is invertible under finite measurements in experiment.

    For simplicity, denote the true value of the determinant of the linear map $\det \Theta_0$ as ${\det}_0$.
    With the Chebyshev's inequality, it can be shown that the probability for the case that the true value of the determinant of the linear map $\Theta$ is not invertible is expressed as
    \begin{equation}
        \mathbb{P}({\det}_0 = 0) \leq \exp\left(-\frac{N}{2\Vert \Theta^{-1} \Vert_2^2}\right).
    \end{equation}
    Therefore, for a given number $N$ of measurements, the linear map $\Theta$ is invertible with the probability $(1 - \delta)$ if the map $\Theta$ satisfies that 
    \begin{equation} \label{eq: invertible}
    \Vert \Theta \Vert_2 \geq \sqrt{\frac{2 D \log \frac{1}{\delta}}{N}},
    \end{equation}
    where $D$ is the dimension of the map $\Theta$.
    For the details of calculation, see Supplementary Note~V.
    \\
    \\\noindent \textbf{Bias of noisy Cancellation}
    \\
    For the noisy cancellation, without mitigating the noises in simulation of inverse noise operation $\mathcal{E}^{-1}$, we would like to estimate the bias of the noisy cancellation $\mathcal{E}_{\lambda}^{-1} \circ \mathcal{E} = \Theta(\mathcal{E}^{-1}) \circ \mathcal{E}$.
    The bias of the expectation of Pauli operator $\hat{O}$ is expressed as
    \begin{align} \label{eq: bias}
        \delta_{\lambda} & = |\mathrm{Tr}[\hat{O}\mathcal{U}(\rho)] - \mathrm{Tr}[\hat{O} \mathcal{E}_{\lambda}^{-1} \circ \mathcal{E} \circ \mathcal{U}(\rho)]| \nonumber\\
        & \leq  p_{\mathcal{Q}}(\mathcal{I} - \mathcal{E}_{\lambda}^{-1} \circ \mathcal{E} ),
    \end{align}
    where we denote $\mathcal{Q} = \mathcal{Q}(A\rightarrow A)$. 
    With $\mathcal{E}_{\lambda}^{-1} = \Theta(\mathcal{E}^{-1})$, we have
    \begin{equation} \label{eq: bound}
        \delta_{\lambda} \leq 2 \Theta_{\lambda} p_{\mathcal{Q}}(\mathcal{E}_{\lambda}^{-1} \circ \mathcal{E}) \leq 2 \Theta_{\lambda} p_{\Theta(\mathcal{Q})}(\mathcal{E}_{\lambda}^{-1}),
    \end{equation}
    where $\Theta_{\lambda} = 1 - \min_i \Theta_{ii}$ is the maximal error probability of the noisy Pauli gates.
    This result allows for estimating the upper bound of the bias in experiments, with the cost of simulation $p_{\Theta(\mathcal{Q})}(\mathcal{E}_{\lambda}^{-1})$ and the calibration of Pauli gates $\Theta_{\lambda}$.
    For the details of the calculation, see Supplementary Note~V.

\begin{figure*}[t]
    \centering
    \includegraphics[width = 0.85\textwidth]{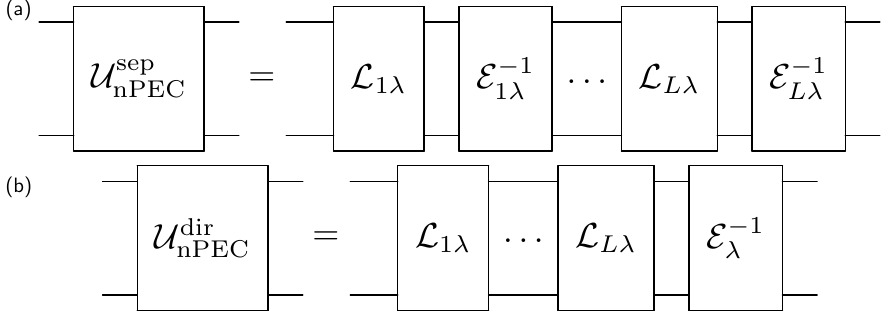}
    \caption{\textbf{Noisy probabilistic error cancellations of multi-layer circuit.}
        \textbf{(a)}, Cancel the errors of the $L$-layer circuit separately in each layer.
        \textbf{(b)}, Cancel the errors of the $L$-layer circuit directly as a whole error of the circuit.}
    \label{fig: layer}
\end{figure*}

    If the circuit consists of $L$ layers of operations $\mathcal{U} = \prod_{i=1}^L \circ \mathcal{L}_i$, where $\mathcal{L}_i$ is the $i$-th layer, there are two strategies to realize the PEC method.
    One is to cancel the error of each layer $\mathcal{E}_i$ separately, and the other is to cancel the total error $\prod_i \mathcal{E}_i$ directly.
    The circuits are shown in Fig.~\ref{fig: layer}.
    Let the noisy realization of circuit be $\mathcal{U}_{\lambda} = \overleftarrow{\bigcirc}_{i = 1}^{L} \mathcal{L}_{i\lambda}$, where $\mathcal{L}_{i\lambda} = \mathcal{E}_i \circ \mathcal{L}_i$, then the error channel of the circuit is 
    \begin{equation}
        \mathcal{E} = \overleftarrow{\bigcirc}_{i = 1}^{L} \mathcal{\tilde{E}}_i,
    \end{equation}
    where $\mathcal{\tilde{E}}_i = ( \overleftarrow{\bigcirc}_{j > i}^{L} \mathcal{L}_{j} ) \circ \mathcal{E}_i \circ (\overrightarrow{\bigcirc}_{j > i}^{L} \mathcal{L}_{j}^{\dagger})$.
    Here, the arrow above the symbol $\bigcirc$ represents the acting direction of layers. 

    We consider the bias of noisy cancellation of the total error of the circuit for these two different strategies.
    For the separate cancellation method, the noisy realization of the error-canceled circuit is 
    \begin{equation}
        \mathcal{U}_{\mathrm{PEC}} = \overleftarrow{\bigcirc}_{i = 1}^{L} \mathcal{L}_{i\mathrm{PEC}},
    \end{equation}
    where $\mathcal{L}_{i\mathrm{PEC}} = \mathcal{E}_{i\lambda}^{-1} \circ \mathcal{E}_i \circ \mathcal{L}_i$.
    The error of the separate cancellation is
    \begin{equation}
        \mathcal{E}_{\mathrm{S}}^{-1} \circ \mathcal{E} = \overleftarrow{\bigcirc}_{i = 1}^{L} \mathcal{\tilde{E}}_{i\mathrm{PEC}},
    \end{equation}
    where $\mathcal{\tilde{E}}_{i\mathrm{PEC}} = ( \overleftarrow{\bigcirc}_{j > i}^{L} \mathcal{L}_{j} ) \circ \mathcal{E}_{i\lambda}^{-1} \circ \mathcal{E}_i \circ (\overrightarrow{\bigcirc}_{j > i}^{L} \mathcal{L}_{j}^{\dagger})$.
    The bias of the noisy separate cancellation is 
    \begin{equation} \label{eq: bias_separate}
        \delta_{\lambda\mathrm{S}} \leq 2 \Theta_{\lambda} \sum_{j=1}^{L} \prod_{i = 1}^{j} p_{\Theta(\mathcal{Q})}(\mathcal{E}_{i\lambda}^{-1}) .
    \end{equation}
    For the detailed calculation, see Supplementary Note~V.
    On the other hand, by applying Eq.~(\ref{eq: bound}), the bias of the direct cancellation method is
    \begin{align} \label{eq: bias_direct}
        \delta_{\lambda \mathrm{D}} & \leq 2 \Theta_{\lambda} p_{\Theta(\mathcal{Q})}(\mathcal{E}_{\lambda}^{-1}\circ \mathcal{E}) \nonumber \\
        & \leq 2 \Theta_{\lambda} \prod_{i = 1}^{L} p_{\Theta(\mathcal{Q})}(\mathcal{E}_{i\lambda}^{-1}).
    \end{align}
    Since the upper bound of the bias of the direct cancellation method, Eq.~(\ref{eq: bias_direct}), is less than the upper bound of the bias of the separate cancellation method, Eq.~(\ref{eq: bias_separate}), the direct cancellation method appears to be more accurate than the separate cancellation method.

    However, there are more results with the detailed analysis of the noisily canceled error $\mathcal{E}_{\lambda}^{-1} \circ \mathcal{E}$.
    We consider a particular case that the noisily canceled error $\mathcal{E}_{\lambda}^{-1} \circ \mathcal{E}$ is a CPTP quantum channel,  $p_{\mathcal{Q}}(\mathcal{E}_{\lambda}^{-1} \circ \mathcal{E}) = 1$.
    For instance, if the noises are uniform on the Pauli gates, $\Theta(\mathcal{E}) = \mathcal{N}\circ \mathcal{E}$, with the commutativity of Pauli diagonal operation, it preserves $\mathcal{E}_{\lambda}^{-1} \circ \mathcal{E} = \mathcal{N} \in \mathcal{Q}(A\rightarrow A)$.
    The bias of the direct cancellation is bounded by
    \begin{equation} \label{eq: bound_direct}
        \delta_{\lambda\mathrm{D}} \leq 2 \Theta_{\lambda},
    \end{equation}
    since $p_{\mathcal{Q}}(\mathcal{E}_{\lambda}^{-1} \circ \mathcal{E}) = 1$.
    For the separate cancellation method, assume the noisily canceled error for each layer $\mathcal{E}_{i\lambda}^{-1} \circ \mathcal{E}_i$ is a CPTP quantum channel; it can be shown that the bias is bounded by
    \begin{equation} \label{eq: bound_separate}
        \delta_{\lambda\mathrm{S}} \leq 2[1 - (1-2\Theta_{\lambda})^{L/2}] \leq 2.
    \end{equation}
    For details of the calculation, see Supplementary Note~V.

\begin{figure*}[t]
    \centering
    \includegraphics[width = \textwidth]{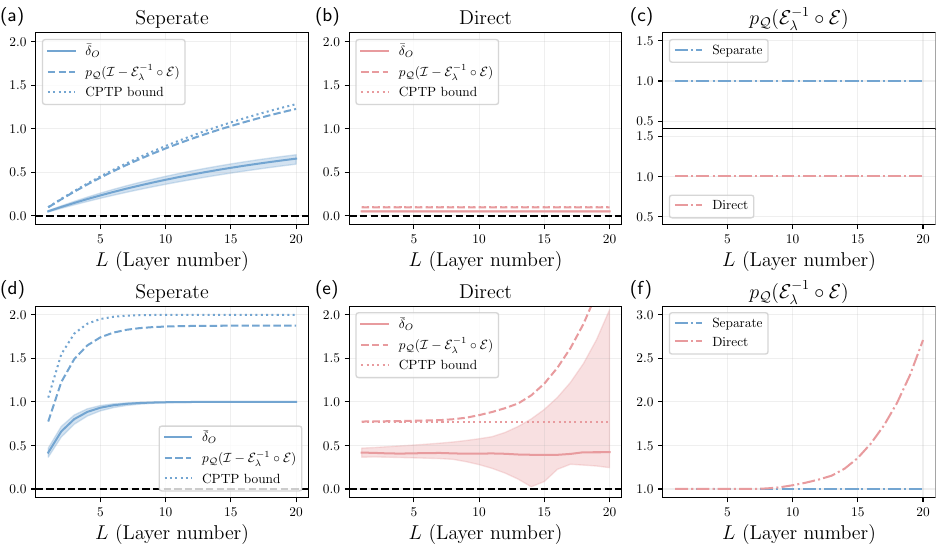}
    \caption{\textbf{Numerical results for the biases of the noisy cancellations with separate and direct methods. }
    \textbf{(a)}, \textbf{(b)}, \textbf{(d)}, and \textbf{(e)}, Colored filling regions denote the biases of expectations for different Pauli operators, and solid curves denote the median of the biases of the operator expectations. 
        Dashed curves denote the distance between the imperfectly canceled error $\mathcal{E}_{\lambda}^{-1} \circ \mathcal{E}$ and $\mathcal{I}$ in the implementability function $p_{\mathcal{Q}}(\mathcal{I}-\mathcal{E}_{\lambda}^{-1} \circ \mathcal{E})$, which upper bounds the bias of the operator expectations by Eq.~(\ref{eq: bias}).
        Dotted curves denote the CPTP upper bound of $p_{\mathcal{Q}}(\mathcal{I}-\mathcal{E}_{\lambda}^{-1} \circ \mathcal{E})$, see Eqs.~(\ref{eq: bound_direct}) and~(\ref{eq: bound_separate}) for the direct and separate cancellation methods, respectively. 
        The error $\mathcal{E}_j$ is assumed to be the same for each layer, $\mathcal{E}_j = \mathcal{E}_0$, and the error $\mathcal{E}_0$ as well as the noises $\mathcal{N}_i$ on Pauli gates $\mathcal{P}_i$ are randomly sampled from the Pauli-Lindblad error model (\ref{eq: Pauli_Lindblad}), with a fixed single-layer error rate $\lambda = \sum_i \lambda_i$. 
        \textbf{(c)} and \textbf{(f)}, Dotted dashed curves denote the implementability function $p_{\mathcal{Q}}(\mathcal{E}_{\lambda}^{-1} \circ \mathcal{E})$ of the imperfectly canceled error for direct and separate cancellation methods.  
        \textbf{(a)}--\textbf{(c)}, For a small error rate $\lambda = 0.05$, since the imperfectly canceled errors $\mathcal{E}_{\lambda}^{-1} \circ \mathcal{E}$  \textbf{(c)}  are CPTP for circuits with a layer number less than $20$, the biases of both the separate \textbf{(a)} and direct \textbf{(b)} cancellation methods are limited by the CPTP upper bound, Eqs.~(\ref{eq: bound_direct}) and~(\ref{eq: bound_separate}), and the bias of the direct cancellation is smaller than the separate cancellation method. 
        \textbf{(d)}--\textbf{(f)}, For a large error rate $\lambda = 0.5$, since the imperfectly canceled error $\mathcal{E}_{\lambda}^{-1} \circ \mathcal{E}$ for the direct cancellation method \textbf{(f)}  will eventually not be CPTP with the cumulation of errors $\mathcal{E} = \mathcal{E}_0^L$. The bias increases exponentially and surpasses the CPTP bound \textbf{(e)}, as the layer number $L$ grows. 
        However, the error $\mathcal{E}_{\lambda}^{-1} \circ \mathcal{E}$ for the separate cancellation method \textbf{(f)} is still CPTP, since each the error of individual layer $\mathcal{E}_{0\lambda}^{-1} \circ \mathcal{E}_0$ is CPTP. The bias of separate cancellation is still bounded by the CPTP upper bound  \textbf{(f)}.  For details of simulation, see Supplementary Note~VI.
    }
    \label{fig: bias}
\end{figure*}

    The numerical results for comparing the two cancellation methods with small and large error rates are shown in Fig.~\ref{fig: bias}.
    When the error rate is sufficiently small, e.g., $\lambda = 0.05$ as used in Fig.~\ref{fig: bias}(a)--(c), the noisily canceled errors for both separate and direct cancellations are CPTP, Fig.~\ref{fig: bias}(c). Thus, the bias is limited by the CPTP upper bounds, as shown in Fig.~\ref{fig: bias}(a) for the separate cancellation Eq.~(\ref{eq: bound_separate}) and Fig.~\ref{fig: bias}(b) for the direct cancellation  Eq.~(\ref{eq: bound_direct}). 
    The bias of direct cancellation is much smaller than the separate cancellation method.
    When the error rate is moderately large, e.g., $\lambda = 0.5$ as used in Fig.~\ref{fig: bias}(d)--(f), the noisily canceled error $\mathcal{E}_{i\lambda}^{-1} \circ \mathcal{E}_i$ of each layer error is still CPTP, thus the noisily canceled error $\prod_i \mathcal{E}_{i\lambda}^{-1} \circ \mathcal{E}_i$ of separate cancellation is also CPTP, as shown in Fig.~\ref{fig: bias}(f), and the bias of separate cancellation is still limited by Eq.~(\ref{eq: bound_separate}), as shown in Fig.~\ref{fig: bias}(d).
    For the direct cancellation method, the noisily canceled error $\mathcal{E}_{\lambda}^{-1} \circ \mathcal{E}$ is CPTP, when the layer is shallow. 
    However, it will no longer be CPTP as the layer increases, as ilustrated in Fig.~\ref{fig: bias}(f).
    The bias of direct cancellation increases exponentially with the layer number accroding to Eq.~(\ref{eq: bias_direct}) in  and will surpass the CPTP upper bound of separate cancellation Eq.~(\ref{eq: bound_separate}), see Fig.~\ref{fig: bias}(e) where the CPTP upper bound is labeled by a dotted curve.
    Thus, for a large error rate, the separate cancellation method is more effective than the direct cancellation method.

    When the error rate is large enough, the noisily canceled error for the individual layer will not be CPTP.
    The bias of both the separate and direct cancellation methods will exponentially increase with the layer number, see Eqs.~(\ref{eq: bias_direct}) and~(\ref{eq: bias_separate}).
    However, for the direct cancellation method, the error of $L$ layers of circuits may not be invertible under a finite measurement precision in the experiment, since its error rate is typically $L$ times the error rate of the one for the individual layer.
    By considering the map $\Xi(\mathcal{N}) = \mathcal{E} \circ \mathcal{N}$, where $\mathcal{E}$ is the error to be canceled, the condition of invertibility of $\mathcal{E}$ is given by Eq.~(\ref{eq: invertible}), as
    \begin{equation}
        \Vert \mathcal{E} \Vert_2 \geq \sqrt{\frac{2 \log \frac{1}{\delta}}{N}}.
    \end{equation}
    In brief, for the case where the circuit is shallow and the error rate is sufficiently small, the direct cancellation method is more accurate.
    If the circuit is deep and the error rate is large, the separate cancellation method is more accurate.

    Intuitively, one may expect a clear criterion for choosing between these two cancellation methods.
    The condition for the criterion requires the solution of a multi-variable polynomial equation $p_{\mathcal{Q}}(\Theta(\mathcal{E}^{-1}) \circ \mathcal{E}) = 1$ in the components $\nu_i$ of $\mathcal{E}$ and the element $\Theta_{ij}$ of $\Theta$.
    However, the solution for the criterion is very complicated, where the error rate $\lambda$ of $\mathcal{E}$ and the maximum error probability $\Theta_{\lambda}$ of the noise map $\Theta$ are not sufficient to determine the criterion.
    In experiments, to verify whether the noisily canceled error is CPTP requires the complete estimation of the noise map $\Theta$, which provides sufficient information for the noiseliess cancellation, thus it need not to consider the bias of noisy cancellation. 
    Therefore, there is no feasible and reliable criterion to distinguish which method, direct or separate, has better performance.
    To quantitatively determine this criterion will be a topic in further investigations.
    Moreover, the error model used for PEC may be inaccurate, which also hinders the verification of the noisily canceled CPTP error. 
    \\
    \\\noindent \textbf{Bias of Inaccurate Error Model}
    \\
    The case where the error model is inaccurate will also induce the bias of the PEC method.
    This is not the main target of this work and has been investigated in other works~\cite{govia2024bounding}. 
    The bias of the PEC from the inaccurate error model is estimated by employing the diamond norm
    \begin{align}
        \delta_O &=|\mathrm{Tr}[O\mathcal{U}(\rho)] - \mathrm{Tr}[O \hat{\mathcal{E}}^{-1} \circ \mathcal{E}\circ \mathcal{U}(\rho)]| \nonumber\\
        & \leq \Vert \mathcal{I} -  \hat{\mathcal{E}}^{-1} \circ \mathcal{E} \Vert_{\diamond} \nonumber\\
        &\leq \sum_{j=1}^L \Vert \mathcal{I} -  \hat{\mathcal{E}}_j^{-1} \circ \mathcal{E}_j\Vert_{\diamond} \prod_{i=1}^{j-1} \Vert \hat{\mathcal{E}}_i^{-1} \circ \mathcal{E}_i\Vert_{\diamond},
    \end{align}
    where $\delta_O = |\braket{O}_{\mathrm{PEC}} - \braket{O}_{\mathrm{exact}}|$ is the bias of the PEC method with the inaccurate error model, $\mathcal{E}_i$ is the exact error channel of $i$-th layer, and $\hat{\mathcal{E}}_i$ is the inaccurate error model learned from experiment.
    With the estimation of the diamond norm of the operation $\mathcal{I} - \hat{\mathcal{E}}_j^{-1} \circ \mathcal{E}_j$, the bias is calculated as
    \begin{equation}
        \delta_O 
        \leq \sum_{j=1}^L \delta_{j} \prod_{i=1}^{j-1} \gamma_i,
    \end{equation}
    where $\gamma_i=p_{\mathcal{Q}}(\hat{\mathcal{E}}_i^{-1})$, 
    \begin{align}
        \delta_{j} & = |1-\nu_0^{(j)}| + \gamma_j - \nu_0^{(j)}, \ \textrm{or} \\
        \delta_{j} & = |1-\nu_0^{(j)}| + T(\{r_k^{(j)}\}),
    \end{align}
    $\nu_0^{(j)} = \frac{1}{4^n} \sum_k r_k^{(i)}$ is the component of $\hat{\mathcal{E}}_j^{-1} \circ \mathcal{E}_j$ in $\mathcal{I}$, $r_k^{(i)} = f_{P_k}^{\mathrm{meas}}/f_{P_k}^{\mathrm{mod}}$ is the ratio of the measured fidelity to the model fidelity of the Pauli gate $P_k$, and 
    \begin{equation}
        T(\{r_k^{(j)}\}) = \frac{4^n-1}{4^n} \sqrt{\sum_k r_k^{(j)}\left(r_k^{(j)} - \frac{1}{4^n-1} \sum_{m\neq k} r_m^{(j)}\right) }.
    \end{equation}

    Since the diamond norm is the implementability function $p_{\mathcal{Q}}$, Eq.~(\ref{eq: diamond}), the bias from the inaccurate error error model is compatible with the bias from noisy cancellation operation estimated in this works.
    The total bias of the noisy PEC with the inaccurate error model is upper bounded as
    \begin{align}
        \delta_{\lambda O} & = |\mathrm{Tr}[O\mathcal{U}(\rho) - \hat{\mathcal{E}}_{\lambda}^{-1} \circ \mathcal{E}\circ \mathcal{U} (\rho)]| \nonumber \\
        & \leq p_{\mathcal{Q}} (\mathcal{I} - \hat{\mathcal{E}}_{\lambda}^{-1}\circ\mathcal{E})
        \leq \delta_O + \delta_{\lambda\mathrm{D},\mathrm{S}}.
    \end{align}

    We next consider the Pauli diagonal error model $\mathcal{E} = \exp{\mathcal{L}(\lambda_i)}$, where $\mathcal{L}(\lambda_i) = \sum_i \lambda_i (\mathcal{P}_i - \mathcal{I})$.
    Assume real error parameters $\{\lambda_i^{(j)}\}$ with deviations from the parameters 
    $\{\hat{\lambda}_i^{(j)}\}$, measured in experiments for the $j$-th layer.
    The mitigated error channel for the $j$-th layer is 
    \begin{equation}
        \hat{\mathcal{E}}_{j}^{-1} \circ \mathcal{E}_{j} = \exp {\mathcal{L}(\Delta \lambda_i^{(j)})},
    \end{equation}
    where $\hat{\mathcal{E}}_{j}$ is the target error channel for the mitigation, $\mathcal{E}_{j}$ is the real error channel, and $\Delta \lambda_i^{(j)} = \lambda_i^{(j)}-\hat{\lambda}_i^{(j)}$. The total mitigated error channel is 
    \begin{equation}
        \hat{\mathcal{E}}^{-1} \circ \mathcal{E}  = \prod_j \hat{\mathcal{E}}_{j}^{-1} \circ \mathcal{E}_{j} = \exp {\mathcal{L}(\Delta \lambda)},
    \end{equation}
    where $\Delta \lambda_i = \sum_j \Delta \lambda_i^{(j)}$.
    The bias of the expectation of the Pauli operator $\hat{O}$ is upper bounded by 
    \begin{align}
        \delta_O & \leq p_{\mathcal{Q}}(\mathcal{I} - \hat{\mathcal{E}}^{-1} \circ \mathcal{E})  \\
        & = p_{\mathcal{Q}}(\hat{\mathcal{E}}^{-1} \circ \mathcal{E}) +|1-\nu_0| - \nu_0. \nonumber
    \end{align}

\begin{figure*}[t]
    \centering
    \includegraphics[width = \textwidth]{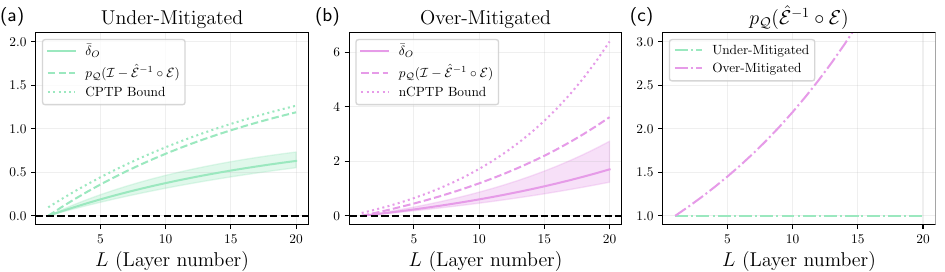}
    \caption{\textbf{Numerical results for the biases of the inaccurate error model with under-mitigated and over-mitigated cases. }
    \textbf{(a)} and \textbf{(b)}, Colored filling regions denote the bias of expectations for different Pauli operators, and  solid curves denote the median of the bias of the operator expectations. 
        Dashed curves denote the distance between the mitigated error $\hat{\mathcal{E}}^{-1} \circ \mathcal{E}$ and $\mathcal{I}$ characterized by the implementability function $p_{\mathcal{Q}}(\mathcal{I}-\hat{\mathcal{E}}^{-1} \circ \mathcal{E})$, which upper bounds the bias of the operator expectations in Eq.~(\ref{eq: bias}).
        Dotted curves denote the CPTP upper bound of $p_{\mathcal{Q}}(\mathcal{I}-\hat{\mathcal{E}}^{-1} \circ \mathcal{E})$, as shown in Eqs.~(\ref{eq: bound_under}) and~(\ref{eq: bound_over}) for the under-mitigated and over-mitigated errors, respectively. 
        The mitigated error channel $\mathcal{\mathcal{E}}_i^{-1} \circ \mathcal{E}_i$ for each layer is assumed to be the same, $\mathcal{E}_i = \mathcal{E}_0$, and randomly sampled from the Pauli-Lindblad error model in 
        Eq.~(\ref{eq: Pauli_Lindblad}), with a fixed single-layer error rate $\Delta\lambda^{\mathrm{under}} = 0.05  \Delta\lambda_i^{\mathrm{under}} \geq 0$ for the under-mitigated error channel \textbf{(a)}, and $\Delta\lambda_i^{\mathrm{over}} = - \Delta\lambda_i^{\mathrm{under}}$ for over-mitigated error channel \textbf{(b)}. \textbf{(c)}, Dotted dashed curves denote the implementability function $p_{\mathcal{Q}}(\hat{\mathcal{E}}^{-1} \circ \mathcal{E})$ for the under-mitigated and over-mitigated errors.  Since the under-mitigated error $\hat{\mathcal{E}}^{-1} \circ \mathcal{E}$ is CPTP, the bias is lower than the CPTP upper bound as shown in \textbf{(a)}. In contrast, the over-mitigated error channel $\hat{\mathcal{E}}^{-1} \circ \mathcal{E}$ is not CPTP, the bias increases exponentially as the layer number $L$ grows, which is bounded by the non-CPTP upper bound in Eq.~(\ref{eq: bound_over})  as shown in \textbf{(b)}. For details of simulation, see Supplementary Note~VI.
    }
    \label{fig: drift}
\end{figure*}

    It is crucial whether the mitigated error $\hat{\mathcal{E}}^{-1} \circ \mathcal{E}$ is CPTP or not, and the error is called under-mitigated or over-mitigated if it is CPTP or not CPTP.
    The numerical results for the bias of under-mitigated and over-mitigated errors are shown in Fig.~\ref{fig: drift}.   
    For the under-mitigated error, $\hat{\lambda}_i^{(j)} \leq \lambda_i^{(j)}$, we have $p_{\mathcal{Q}}(\hat{\mathcal{E}}^{-1} \circ \mathcal{E}) = 1$, as shown in Fig.~\ref{fig: drift}(c), and the upper bound is shown in Fig.~\ref{fig: drift}(a) as
    \begin{equation} \label{eq: bound_under}
        \delta_{O} \leq 2\left[1 -\mathrm{e}^{-\Delta\lambda}\right],
    \end{equation}
    where $\Delta\lambda = \sum_i \Delta\lambda_i > 0$.
    For the over-mitigated error, $\hat{\mathcal{E}}^{-1} \circ \mathcal{E}$ is not CPTP as illustrated in Fig.~\ref{fig: drift}(c), the upper bound increases as the layer number increases, as shown in Fig.~\ref{fig: drift}(b)
    \begin{equation} \label{eq: bound_over}
        \delta_{O} \leq \mathrm{e}^{2 \Delta\lambda_{-}} - 2 \mathrm{e}^{-\max\{\Delta\lambda,0\}} + 1,
    \end{equation}  
    where $\Delta \lambda_{-} \equiv \sum_i |\min\{\Delta\lambda_i,0\}|$.
    Therefore, when considering the error model violation, the under-mitigated error channel outperforms the under-mitigated error channel. 

    For the error beyond the description with the Pauli diagonal error model, it can be randomly compiled into the Pauli diagonal error model by using the well-performed Pauli twirling technique.
    However, when the number $N$ of shots for Pauli twirling is not sufficiently large, the inaccurate error model will yield a deviation from the estimation.
    This deviation will be suppressed with the increase of $N$ as $\sim 1/\sqrt{N}$ according to the large number theorem, so it can be interpreted as a statistical error of the estimation.
    Given a tolerant precision $\delta$ of the estimation and a tolerant probability $\gamma \ll 1$ for the failure, the number of shots $N$ should satisfies
    \begin{equation}
        N \gtrsim \frac{L^2}{\delta^2} (A - B \log\gamma),
    \end{equation}
    where $A, B \geq 0$ are constants for the specific error model.
    More detailed discussions can be found in Supplementary Note~V.           
    \\
    \\\textbf{\Large{Discussion}}
    \\
    The implementability function is related to several quantities in quantum information theory and has further applications in quantum information processing.
    Here, we discuss the relationship between the implementability function and the diamond norm, logarithmic negativity, as well as purity.
    These results can reduce to the results about CPTP operations $\mathcal{Q}(A\rightarrow A)$ in other works.
    However, our technique does not stem from the semidefinite property of $\mathcal{Q}$ but uses the convexity and the free property of resource theories and can be applied to a general case.

    We first consider the relationship between the implementability function and the diamond norm~\cite{kitaev1997quantum,watrous2018theory}.
    For a system $A$, given a convex free set of states $\mathcal{F}(A)$, the implementability function $p_{\mathcal{F}(A)}$ can be defined.
    Naturally, it induces a norm on the space of quantum operations $\mathcal{B}(A \rightarrow A)$
    \begin{equation}
        \Vert \mathcal{N}\Vert_{\diamond \mathcal{F}(A)} \equiv \max \frac{p_{\mathcal{F}(A)}(\mathcal{N}(\rho))}{p_{\mathcal{F}(A)}(\rho)}.
    \end{equation}
    This norm is a generalization of the diamond norm
    \begin{equation}
        \Vert \mathcal{N}\Vert_{\diamond} \equiv \max_{\rho \in \mathcal{Q}(A \otimes A)} \frac{\Vert \mathrm{id} \otimes \mathcal{N}(\rho)\Vert_1}{\Vert \rho\Vert_1},
    \end{equation}
    where the free set is $\mathcal{Q}(A \otimes A)$, since the implementability function $p_{Q(A)} = \Vert \cdot \Vert_1$, with respect to the set of all quantum state $\mathcal{Q}(A)$ is the trace norm $\Vert \cdot \Vert_1$ (see Supplementary Note~VII).
    The norm $\Vert \cdot \Vert_{\diamond \mathcal{F}(A)}$ is the Minkowski functional of the set $\mathcal{F}_{\max}(A \rightarrow A)$ of resource non-generating (RNG) operations~\cite{RevModPhys.91.025001}, i.e. the maximal assignment of operation for the free set $\mathcal{F}(A)$ of state (see Supplementary Note~VII for the proof)
    \begin{equation} \label{eq: diamond}
        \Vert \mathcal{N}\Vert_{\diamond \mathcal{F}(A)} = p_{\mathcal{F}_{\max}(A \rightarrow A)}(\mathcal{N}).
    \end{equation}
    Therefore, the diamond is the exponential of the physical implementability 
    \begin{equation} 
        \Vert \mathcal{N}\Vert_{\diamond} = p_{\mathcal{Q}(A\rightarrow A)}(\mathcal{N}).
    \end{equation}
    This result has also been proved with semidefinite programs~\cite{regula2021operational}, where $p_{\mathcal{Q}(A \rightarrow A)}$ is denoted as $\Vert \cdot \Vert_{\blacklozenge}$. 
    We give a different proof (see Supplementary Note~VII) without semidefinite programs, which is simpler.

    In addition, we consider the relationship between the implementability function and the logarithmic negativity.
    The following results generalize the results of physical implementability~\cite{guo2023noise}.
    The logarithmic negativity is defined as~\cite{PhysRevA.65.032314,PhysRevLett.95.090503}
    \begin{equation}
        E_N(\rho) = \log \Vert \rho^{T_B} \Vert_1,
    \end{equation}
    which denotes the negativity of the partial transposed density matrix $\rho^{T_B}$, and is effective to detect entanglement~\cite{PhysRevLett.77.1413,HORODECKI19961}.
    Since $p_{\mathcal{Q}(AB)} = \Vert \cdot \Vert_1$, it is the logarithm of implementability function
    \begin{equation}
        E_N(\rho) = \log p_{T_B(\mathcal{Q})(AB)}(\rho)
    \end{equation}
    on the free set $T_B(\mathcal{Q})(AB)$ of partial transpose of quantum states, since the partial transpose $T_B(\rho) = \rho^{T_B}$ is a linear involution $T_B^2 = \mathrm{id}_B$.
    Given a convex free set $\mathcal{F}(AB)$, for $\rho \in \mathit{B}(AB)$, we can define a generalized version $E_{N \mathcal{F}(AB)}$ of logarithmic negativity with respect to a convex set $\mathcal{F}(AB)$ as
    \begin{equation}
        E_{N \mathcal{F}}(\rho) = \log p_{T_B(\mathcal{F})(AB)}(\rho),
    \end{equation}
    where $T_B(\mathcal{F})(AB)$ is the partial transpose of the free set $\mathcal{F}(AB)$.
    With the so-defined generalized logarithmic negativity and the sub-multiplicity of implementability function, it follows that for $\rho = \mathcal{N}(\rho_0)$, where $\rho_0 \in \mathcal{B}(AB)$, $\mathcal{N} \in \mathcal{B}(AB \rightarrow AB)$
    \begin{align}
        E_{N \mathcal{F}(AB)}(\rho) & - E_{N \mathcal{F}(AB)}(\rho_0) \nonumber\\
        & \leq \log p_{T_B(\mathcal{F})(AB \rightarrow AB)}(\mathcal{N}), 
    \end{align}
    where the action of partial transpose on operations is induced as
    \begin{equation}
        T_B(\mathcal{N}) = T_B \circ \mathcal{N} \circ T_B .
    \end{equation}
    For the proof, see Supplementary Note~VII.
    This inequality is tight when 
    \begin{equation}
        p_{T_B(\mathcal{F})(AB \rightarrow AB)}(\mathcal{N}) = \Vert \mathcal{N} \Vert_{\diamond T_B(\mathcal{F})(AB)},
    \end{equation}
    and by Eq.~(\ref{eq: diamond}), i.e., the free set of operations is a resource nongenerating set
    \begin{equation}
        \mathcal{F}(AB \rightarrow AB) = \mathcal{F}_{\max}(AB \rightarrow AB).
    \end{equation} 
    When the operation $\mathcal{N}$ is invariant under the partial transpose, namely
    \begin{equation}
        T_B \circ \mathcal{N} \circ T_B = \mathcal{N},
    \end{equation}
    the inequality reduces to
    \begin{equation}
        E_{N \mathcal{F}(AB)}(\rho) - E_{N \mathcal{F}(AB)}(\rho_0) \leq \log p_{\mathcal{F}(AB \rightarrow AB)}(\mathcal{N}).
    \end{equation}
    In particular, it holds for the local operations $\mathcal{N} = \sum_i q_i \mathcal{N}_i^{A} \otimes \mathcal{N}_i^{B}$, which is invariant under the partial transpose.

    Purity is the square of the Frobenius norm $\Vert A \Vert_2 = \sqrt{\mathrm{Tr} (A^{\dagger} A)}$.
    Let $\sigma \in \mathcal{B}(A)$, $\mathcal{N} \in \mathcal{B}(A \rightarrow A)$.
    For the Frobenius norm, we have
    \begin{equation}
        \frac{\Vert \mathcal{N}(\sigma) \Vert_2}{\Vert \sigma \Vert_2} \leq p_{\mathcal{F}(A)}(\mathcal{N}) \max_{l \in A^+ \cup A^-} \Vert \mathcal{F}_l \Vert_F,
    \end{equation}
    where $\Vert \cdot \Vert_F$ is an induced norm of the Frobenius norm 
    \begin{equation}
        \Vert \mathcal{M} \Vert_F = \max_{i} \Vert \mathcal{M}_i \Vert_F = \max_{i} \max_{\sigma_i} \frac{\Vert \mathcal{M}_i(\sigma_i) \Vert_2}{\Vert \sigma_i \Vert_2},
    \end{equation}
    where $\mathcal{M}_i$ is the Jordan decomposition of the operation $\mathcal{M}$ as the matrix acting on the vector space of state $\mathcal{B}(A)$.
    Here, $\mathcal{F}_l$ is the extreme points of the free set, where $\mathcal{N}$ has optimal decomposition on them.
    In particular, if the operation $\mathcal{N}$ is identity, i.e., $\mathcal{N}(I) = I$, denoting $\sigma = \rho - I/D$, we have
    \begin{align}
        \log \frac{P(\mathcal{N}(\rho)) D - 1}{P(\rho) D - 1} & \leq 2 \log p_{\mathcal{F}(A)}(\mathcal{N}) \nonumber \\
        &~~~~+ 2 \log \max_{l \in A^+ \cup A^-} \Vert \mathcal{F}_l \Vert_F.
    \end{align}
    Moreover, if $\mathcal{N} = \sum_l x_l \mathcal{U}_l$ is a unitary mixed operation, where the unitaries $\mathcal{U}_l \in \mathcal{F}(A\rightarrow A)$ are free, namely $\mathcal{F}_l = \mathcal{U}_l$, we have
    \begin{equation}
        \log \frac{P(\mathcal{N}(\rho)) D - 1}{P(\rho) D - 1} \leq 2 \log p_{\mathcal{F}(A)}(\mathcal{N}).
    \end{equation}    
    For the proof, see Supplementary Note~VII.
    \\
    \\
    \noindent \textbf{\Large{Conclusion}}
    \\
    The implementability function $p_{\mathcal{F}}$ is the generalization of physical implementability to arbitrary convex free set $\mathcal{F}$, which evaluates the minimal cost to simulate a resource with the free resources by quasiprobability decomposition.
    We investigate the noisy realization of the PEC method based on the properties of the implementability function. 
    We demonstrate the way to optimally simulate the inverse operation of the error channel with the noisy Pauli basis.
    This method requires that the noisy Pauli gates can be reversed into ideal Pauli gates.
    We give the condition under which the invertibility is guaranteed with a tolerant probability of failing under finite measurements in experiments.
    We also derive upper bounds of the bias of the noisy cancellation of error channel without canceling the noise on Pauli gates.
    It shows that to mitigate the error channels in circuits with many layers, canceling the total error directly has better performance than canceling the layer errors separately when the error rate and layer number are small.
    Otherwise, the separate cancellation has better performance. 

    Moreover, we discuss several quantum-information quantities that are relevant to the implementability function, including the diamond norm, logarithm negativity, and purity.
    We propose a norm on quantum operations based on a given implementability function of state, which is a generalization of the diamond norm.
    This norm denotes the implementability function of the resource non-generating set of operation.
    It directly implies that the diamond norm is the exponential of physical implementability, which has also been proved with semidefinite programs~\cite{regula2021operational}.
    Moreover, we derive the relation between the implementability function and the logarithmic negativity and purity, which extends the existed results~\cite{guo2023noise}. 

    Our results are mainly based on the convexity of the free set.
    However, the composition of operation may imply more properties for quantifying resources, which may be left to further research.  
    We hope that our results will be useful for the probabilistic error cancellation method in practice.
    We also expect that the implementability function will have more applications in the field of quantum information processing.  
    \\
    \\
    \noindent \textbf{\Large{Methods}}
    \\
    Let $A, B, C, \dots$ denote the systems; the Hilbert spaces of these systems are $\mathcal{H}_{A}, \mathcal{H}_{B}, \mathcal{H}_{C}$.
    The set of bounded linear operators on the Hilbert space $\mathcal{H}_{A}$ is denoted as $\mathcal{B}(A)$, to which the density matrices of $A$ system belong.
    The linear operation on the operators from $\mathcal{B}(A)$ to $\mathcal{B}(B)$ is denoted as $\mathcal{B}(A \rightarrow B)$.

    The quantum resources form a convex set $\mathcal{Q}\subset \mathcal{B}/\mathbb{R}$ of space of normalized resources.
    Here, we focus on the quantum channels $\mathcal{Q}(A \rightarrow A)$, which are the completely positive and trace preserving (CPTP) quantum operations.
    For the noisy cancellation, the implementable channels can be expressed as $\Theta(\mathcal{Q})(A \rightarrow A)$, where $\Theta$ is the linear map from ideal channels to noisy channels.
    In this section, the symbol $(A\rightarrow A)$ for quantum operation will be omitted for simplicity.

    The quasiprobability decomposition technique used in the PEC has also been used in other quantum information processing tasks, such as entanglement forging~\cite{PRXQuantum.3.010309}, where the entangled states are simulated with separable states, and circuit knitting~\cite{Mitarai_2021,Mitarai2021overhead,10.1109/TIT.2023.3310797}, where the nonlocal operations are simulated with local operations.
    Therefore, it is worth extending the definition of the physical implementability to any convex free set of quantum resources.
    (see Supplementary Note~I for a brief introduction to the framework of quantum resource theories).
    In general, for a given convex set $\mathcal{F}$ of  freely implementable resources, the implementability function can be defined as
    \begin{equation}
    p(N) = \inf_{\mathcal{F}} \sum_i |x_i|, 
    \end{equation} 
    over all decomposition $N = \sum_i x_i E_i$ with $E_i \in \mathcal{F}$.
    Our definition does not take the logarithm, since the implementability function $p(N)$ so-defined is just the minimal number of free operations used to simulate a resourceful operation with quasiprobability decomposition technique.
    Meanwhile, this quantity is also related to the robustness measure $R$ in the quantum resource theories~\cite{RevModPhys.91.025001,PhysRevLett.115.070503,PhysRevA.59.141,PhysRevResearch.3.033178} as
    \begin{equation}
        p = 2 R + 1.
    \end{equation}
    The robustness depicts the minimal amount of mixing with free resources to wash out all resources~\cite {PhysRevA.59.141}.
    The implementability function operationally corresponds to the inverse problem of the robustness.

    The infimum of the cost is attained if the convex set $\mathcal{F}$ is bounded closed
    \begin{equation} \label{eq: min}
        p(N) = \min_{\mathcal{F}} \sum_i |x_i| < \infty.
    \end{equation} 
    Hereafter, we assume that the free set $\mathcal{F}$ is a bounded closed convex set.
    Then, the implementability function has the following basic properties:\\
    \begin{enumerate}[i)]
    \item[(i)] Faithfulness
    \begin{equation} \label{eq: faithfulness}
        p(N) \geq 1, \ \mathrm{and} \ N \in \mathcal{F}, \ \mathrm{iff} \ p(N) = 1.
    \end{equation}
    \item[(ii)] Sub-linearity
    \begin{equation} \label{eq: sub-linearity}
        p(a N_1 + b N_2) \leq |a| p(N_1) + |b| p(N_2).
    \end{equation}
    \item[(iii)] Composition sub-multiplicity
    \begin{equation}\label{eq: sub-multiplicity}
        p(M \circ N) \leq p(M) p(N).
    \end{equation}
    \item[(iv)] Tensor-product sub-multiplicity
    \begin{equation} \label{eq: tensor-product}
        p(M \otimes N) \leq p(M) p(N).
    \end{equation}
    \item[(v)] If $\mathcal{F}_1 \subset \mathcal{F}_2$, 
    \begin{equation} \label{eq: subset}
        p_{\mathcal{F}_1} \geq p_{\mathcal{F}_2}.
    \end{equation}
    \end{enumerate}

    In general, not all the resources can be implemented by the resources that can be freely implemented with quasiprobability decomposition.  
    One example is that if the freely implementable resources are local operations and classical channels (LOCC), the CNOT gate is not implementable~\cite{guo2023noise}.
    Therefore, we consider the affine space $\mathcal{A} = \mathrm{aff}(\mathcal{F})$, and the vector space $V = \langle \mathcal{F} \rangle$ generated by the free set $\mathcal{F}$.
    Then, the implementability function $p_{\mathcal{F}}$ is the Minkowski gauge function~\cite{osborne2014locally} of the convex set $\mathcal{C} = \mathrm{conv}(\mathcal{F} \cup -\mathcal{F})$, when constrained on the affine space $\mathcal{A} = \mathrm{aff}(\mathcal{F})$, as
    \begin{equation} \label{eq: geometry}
        p_{\mathcal{F}} = p_{\mathcal{C}}|_{\mathcal{A}} \equiv \inf\{\alpha \geq 0: N \in \alpha \mathcal{C}\}.
    \end{equation}
    This implies that the implementability function is a ratio of the length of lines, which is an affine invariant preserved by affine transformations $\mathrm{GA}(V)$~\cite{prasolov2001geometry}.
    In the vector space $V$, the implementability function is a norm, which is the so-called base norm for the free set $\mathcal{F}$~\cite{hartkamper1974foundations,regula2017convex}.
    Actually, for any norm $\Vert\cdot \Vert$, since its sub linearity, there is a balanced convex set $\mathcal{C}_{\Vert\cdot \Vert} = \{N: \Vert N \Vert \leq 1\}$,
    such that the norm is the Minkowski gauge function of $\mathcal{C}_{\Vert\cdot \Vert}$.
    Therefore, the implementability function may be potentially related to many norms widely used in quantum information theory. 
    For the proofs of the properties above, see Supplementary Note~II. 
    For more mathematical properties and structures of the implementability function, see Supplementary Note~III and Supplementary Note~~IV. 
    \\
    \\
    \\
    \\
    \textbf{\Large{Acknowledgments}}\\
    This work was supported by the Innovation Program for Quantum Science and Technology (Grant No.~2021ZD0301800), the National Natural Science Foundation of China (Grants Nos.~T2121001, 92265207, 12122504, 12475017), the Natural Science Foundation of Guangdong Province (Grant No.~2024A1515010398). We also acknowledge the supported from the Synergetic Extreme Condition User Facility (SECUF) in Beijing, China.
    \\
    \\
    \\
    \\
    \textbf{\Large{Author contributions}}\\
    H.F and Y.-R.Z. supervised the project; T.-R.J. proposed the idea; T.-R.J. performed the numerical simulations and discussed with Y.-R.Z. and K.X.; T.-R.J., Y.-R.Z., and H.F co-wrote the manuscript, and all authors contributed to the discussions of the results and development of the manuscript.
    \\
    \\
    \\
    \textbf{\Large{Competing interests}}\\
    The authors declare no competing interests.
    \\
    \\
    \\
    \textbf{\Large{Data availability}}\\
    The datasets generated in this study have been deposited in the Figshare repository~\cite{Jin2025}:\\ \href{https://doi.org/10.6084/m9.figshare.29290544}{https://doi.org/10.6084/m9.figshare.29290544}.
    \\
    \\
    \\
    \textbf{\Large{Code availability}}\\
    The codes used in numerical simulation have been deposited in the Figshare repository~\cite{Jin2025}:\\ \href{https://doi.org/10.6084/m9.figshare.29290544}{https://doi.org/10.6084/m9.figshare.29290544}.

%

~\\

\end{document}


\title{Supplementary Information for ``Noisy Probabilistic Error Cancellation and Generalized Physical Implementability''}

\author{Tian-Ren Jin}
\affiliation{Institute of Physics, Chinese Academy of Sciences, Beijing 100190, China}
\affiliation{School of Physical Sciences, University of Chinese Academy of Sciences, Beijing 100049, China}

\author{Yu-Ran Zhang}
\email{yuranzhang@scut.edu.cn}
\affiliation{School of Physics and Optoelectronics, South China University of Technology, Guangzhou 510640, China}

\author{Kai Xu}
\affiliation{Institute of Physics, Chinese Academy of Sciences, Beijing 100190, China}
\affiliation{School of Physical Sciences, University of Chinese Academy of Sciences, Beijing 100049, China}
\affiliation{Beijing Academy of Quantum Information Sciences, Beijing 100193, China}
\affiliation{Hefei National Laboratory, Hefei 230088, China}
\affiliation{Songshan Lake Materials Laboratory, Dongguan 523808, China}
\affiliation{CAS Center for Excellence in Topological Quantum Computation, UCAS, Beijing 100190, China}

\author{Heng Fan}
\email{hfan@iphy.ac.cn}
\affiliation{Institute of Physics, Chinese Academy of Sciences, Beijing 100190, China}
\affiliation{School of Physical Sciences, University of Chinese Academy of Sciences, Beijing 100049, China}
\affiliation{Beijing Academy of Quantum Information Sciences, Beijing 100193, China}
\affiliation{Hefei National Laboratory, Hefei 230088, China}
\affiliation{Songshan Lake Materials Laboratory, Dongguan 523808, China}
\affiliation{CAS Center for Excellence in Topological Quantum Computation, UCAS, Beijing 100190, China}

\maketitle
\beginsupplement
\tableofcontents
\newpage

\section{Brief Introduction to 
Quantum Resource Theories} \label{app: QRT}

The most of the physics theories are devoted to describe the rules of nature, and it seems that human's factors should be avoid in the physics theories. 
However, thermodynamics, especially the second law of thermodynamics, implies the preciousness of works, thus contains the economics perspective. 
In the studies of entanglement, many similarities to thermodynamics have been found~\cite{PhysRevLett.80.5239,PhysRevLett.89.240403}.
Moreover, entanglement, athermality, and asymmetry have been recognized as the resources for important task in quantum information processing, thermodynamics, and more areas~\cite{doi:10.1142/S0217979213450197,PhysRevLett.115.070503,PhysRevLett.118.060502,Sparaciari2020firstlawofgeneral}.

Quantum resource theories (QRTs) take the economics perspective into physics, and phrase the scarcity of resources in an abstract, axiomatic framework~\cite{LIEB19991,lieb2013entropy,PhysRevLett.117.260601,FRITZ_2017,COECKE201659}.
It has been formulated as a symmetric monoidal category~\cite{FRITZ_2017,COECKE201659}.
Different from the physics theories of nature, QRTs offer a framework to depict the operations in laboratory.
In this Appendix, we will briefly introduce some basic structures of QRTs~\cite{RevModPhys.91.025001}, which may be useful in this work.

Let the $A, B, C, \dots$ denote some systems, the Hilbert spaces of these systems are $\mathcal{H}_{A}, \mathcal{H}_{B}, \mathcal{H}_{C}$.
The set of bounded linear operators on the Hilbert space $\mathcal{H}_{A}$ is denoted as $\mathcal{B}(A)$, where the density matrices of $A$ system belong.
The linear operation on the operators from $\mathcal{B}(A)$ to $\mathcal{B}(B)$ is denoted as $\mathcal{B}(A \rightarrow B)$.
We use $X$ denote a variable of system labels, like $X = A \equiv \mathbb{C} \rightarrow A$ for state operators or $X = A \rightarrow B$ for operations. 
For simplicity, we may omit the system label variable $X$ if not necessary.

The density matrix $\rho$ of a quantum system $A$ should satisfy the properties of Hermitian $\rho^{\dagger} = \rho$, normalization $\mathrm{Tr}\rho = 1$ and semi-definite positivity $\rho \geq 0$.
Correspondingly, an operation $\mathcal{E}$ preserves the properties of Hermitian, normalization, and semi-definite positivity of density matrix are called Hermitian-preserving (HP), trace-preserving (TP), and positive (P).
Moreover, if the operation $\mathcal{E} \in \mathcal{B}(A\rightarrow A)$ has the property that $\mathcal{E} \otimes \mathrm{id}_B$ is also positive with additional system $B$, $\dim B = k$, where $\mathrm{id}^B$ is the identical operation on $B$ that maps the density matrices of $B$ to themselves, it is called as $k$-positive.
In particular, if the operation $\mathcal{E}$ is $\dim A$-positive, it is also positive for $k>\dim A$, which is called as completely positive (CP).
The quantum resource is a map $\mathcal{Q}: X \mapsto \mathcal{Q}(X)$, where $\mathcal{Q}(X) \subset \mathcal{B}(X)$ is the set of all physical resource of quantum theory.
For static resource $X = A$, $\mathcal{Q}(A)$ is the set of all density matrix $\rho$ of system $A$.
For dynamical resource $X = A \rightarrow B$, $\mathcal{Q}(A \rightarrow B) = \textrm{CPTP}$ is the set of all CPTP channels from system $A$ to $B$.
When the system label variable $X$ is default, we do not distinguish the map $\mathcal{Q}$ with its image $\mathcal{Q}(X)$, so do other maps.

The set of resource $\mathcal{B}$ have the structure of algebra.
The addition $+$ represents the combination of resource.
The scalar multiplication over $\mathbb{R}$ can be induced by the addition with limit.
The multiplication $\circ$ represents the composition of the resource.
The normalized resources $\mathcal{T} = \mathcal{B}/\mathbb{R}$ is a map to affine space, since if $N_1, N_2 \in \mathcal{T}(X)$, then $(a N_1 + b N_2) \in \mathcal{T}(X)$ for $a + b = 1$, while the quantum resource $\mathcal{Q}(X)$ is convex in $\mathcal{T}(X)$, which requires $a, b \geq 0$ in addition.
Here, if $X = A$, the resource $N_i = \rho_i$ are quasi-density matrix, while $X = A \rightarrow B$, the resource $N_i = \mathcal{N}_i^{A \rightarrow B}$ are operations.
The convexity represents the classically probabilistic mixing.

To assign the resources in laboratory, it equivalence to depict the objects out of resource.
The objects available in laboratory is called as free to a quantum resource theory.
In general, the free set can be interpreted as a map $\mathcal{F}: X \mapsto \mathcal{F}(X) \subset \mathcal{B}(X)$ from the systems to the sets of free states or operation of the systems.
To consistent with the scarcity of resource, the free set should satisfy the conditions: 
\begin{itemize}
    \item [(i)] $\mathrm{id}^A \in \mathcal{F}(A \rightarrow A)$.
    \item [(ii)] If $\mathcal{N} \in \mathcal{F}(A \rightarrow B), \mathcal{M} \in \mathcal{F}(B \rightarrow C)$, then $\mathcal{N} \circ \mathcal{M} \in \mathcal{F}(B \rightarrow C)$.
\end{itemize}
Moreover, if we are going to process more than one system simultaneously, the free set should admit a tensor-product structure:
\begin{itemize}
    \item [(i)] Completely free: if $\mathcal{N} \in \mathcal{F}(A \rightarrow B)$, then $\mathcal{N} \otimes \mathrm{id}_C \in \mathcal{F}(AC \rightarrow BC)$.
    \item [(ii)] If $\rho \in \mathcal{F}(B)$, then the appending operation $\Phi_{\rho}(\sigma) \equiv \sigma \otimes \rho$ is free, $\Phi_{\rho} \in \mathcal{F}(A \rightarrow AB)$.
    \item [(iii)] Discarding a system is free, $\mathcal{F}(A \rightarrow \mathbb{R}) \neq 0$.
\end{itemize}
Note that the uniqueness of the discarding operation is called as terminality of the tensor unit, which is equivalent to satisfying causality~\cite{coecke2013causal,coecke2016terminality,chiribella2016quantum}.

Given a free set of states $\mathcal{F}(A\rightarrow A)$, based on the ``free'' properties of the free set, the free set of operations $\mathcal{F}(A\rightarrow A)\subset  \mathcal{F}_{\max}(A \rightarrow A)$ is confined by the set of resource non-generating (RNG) operations 
\begin{equation}
    \mathcal{F}_{\max}(A \rightarrow A) = \{\mathcal{N}: \mathcal{N}(\rho) \mathcal{F}(A) \forall \rho \in \mathcal{F}(A)\}.
\end{equation}
Moreover, if the free set is admit the tensor product, then the RNG property can be extended to $k$-RNG and completely-RNG like the positive and free properties.  

Except for the free property introduced in above, the free set may have other properties, such as convexity, affine, and more.
With these additional property, the QRTs will have more structures.
In this work, we only consider the convex QRT.

\section{Properties of implementability function} \label{app: property}

\subsection{Minimum}

\begin{definition} \label{def: implementability}
    The implementability function of element $N \in \mathcal{A}$ with respect to the free set $\mathcal{F}$ is
    \begin{equation}
        p(N) = \inf_{\mathcal{F}} \sum_i |x_i|, 
    \end{equation} 
    over all $E_i \in \mathcal{F}$ that $N = \sum_i x_i E_i$.    
\end{definition}\noindent
The implementability function of element $N$ is defined as the infimum of the cost of its quasiprobability decomposition.
Intuitively, there is no doubt that the infimum $p(N)$ is attained when the free set is closed (and bounded).
In Ref.~\cite{Jiang2021physical}, the proof is skipped with a comment that it can be proved by some followed results, which, however, actually assume the infimum is attained.
Although the intuition is correct, we would like to prove it explicitly.
\begin{theorem} \label{the: min}
    If the convex free set $\mathcal{F}$ is bounded closed, then for all element $N$, there exists decomposition $N = \sum x_i E_i$ attaining the infimum, the implementability function
    \begin{equation}
        p(N) = \sum_i |x_i| < \infty.
    \end{equation}
\end{theorem}\noindent
To prove this, we need a lemma
\begin{lemma} \label{lem: extreme}
    Let $\mathrm{ext}(\mathcal{F}) = \{F_l\}$ be the set of all extreme points of $\mathcal{F}$, then 
    \begin{equation}
        p(N) = \inf_{\{F_l\}} \sum_l |n_l|,
    \end{equation}
    where $N = \sum_l n_l F_l$.
\end{lemma}
\begin{proof}
    Since $\mathcal{F}$ is bounded closed, it is the convex closure of its extreme points $\mathrm{ext}(\mathcal{F})=\{F_l\} \subset \mathcal{F}$ by Minkowski's theorem~\cite{prasolov2001geometry,hug2020lectures}, and $\mathcal{A}$ is the affine closure of $\mathrm{ext}(\mathcal{F})$, thus $p(N) \leq \inf_{\{F_l\}} \sum_l |n_l|$.
    On the other hand, $\forall E_i \in \mathcal{F}$, there exist convex combinations 
    \begin{equation}
        E_i = \sum_l e_{il} F_l, \quad \sum_l e_{il} = 1,\ e_{il} \geq 0.
    \end{equation}
    Therefore, For arbitrary decomposition 
    \begin{equation}
        N = \sum_i x_i E_i = \sum_{l} n_l F_l,
    \end{equation}
    there exist $n_l = \sum_i x_i e_{il}$.
    Then, 
    \begin{equation}
        \inf_{\{F_i\}} \sum_l |n_l| \leq \sum_l |n_l| \leq \sum_l \sum_i |x_i| e_{il} = \sum_{i} |x_i|,
    \end{equation}
    for all possible $\{x_i\}$, and 
    \begin{equation}
        \inf_{\{F_i\}} \sum_l |n_l| \leq p(N).
    \end{equation}
\end{proof}
\begin{proof}[Proof of Theorem~\ref{the: min}]
    First, we prove the infimum exists.
    Since $\mathcal{A} = \mathrm{aff}(\mathcal{F})$, let $n = \dim \mathcal{A} = \dim \mathcal{F}$.
    Then, there exist $n+1$ affinely independent elements $\{E_i\} \in \mathcal{F}$ that 
    \begin{equation}
        N = \sum_i x_i E_i.
    \end{equation}
    Therefore, the set $\left\{\sum_i |x_i|\right\}$ is not empty, and the infimum $p(N) \leq \sum_i |x_i| \leq (n+1) \max_i |x_i| \leq \infty$ exists.

    By Lemma~\ref{lem: extreme}, $p(N) = \inf_{\{F_l\}} \sum_l |n_l|$. 
    If the extreme points $\mathrm{ext}(\mathcal{F})$ are affine independent, then the decomposition of $N$ into $F_l$ is unique, which is what we want.
    Otherwise, there are different linear constraints of $F_l$ satisfied
    \begin{equation}
        C_s = \sum_l f_{sl} F_l = 0,
    \end{equation}
    where $s \in S$ is some index.
    Then, all the decompositions of $N$ into $F_l$
    \begin{equation}
        N = \sum_s x_s C_s + N_0,
    \end{equation}
    where $N_0$ is some decomposition of $N$, are isomorphic to the space $\{(x_s)\} = \mathbb{R}^{s}$.
    Since $p(N)$ exists, there is a converged sequence of 
    \begin{equation}
        p_k = g[(x_{ks})] = \sum_{l} \left|\sum_s x_{ks} f_{sl} + n_{0l}\right| \in \left\{\sum_i |x_i|\right\}
    \end{equation}
    Moreover, the pre-implementability function $g$ of decompositions is continuous, thus the sequence $(x_{ks}) \in \mathbb{R}^{s}$ is also convergent.
    Since $\mathbb{R}^{s}$ is close, there is a point $(x^*_s) \in \mathbb{R}^{s}$ such that $g[(x^*_s)] = p(N)$, and the corresponding decomposition
    \begin{equation}
        N = \sum_s x_s^* C_s + N_0
    \end{equation}
    is what we want.
\end{proof}\noindent
With this Theorem~\ref{the: min}, Lemma~\ref{lem: extreme} can be restricted:
\begin{corollary}[extreme-point decomposition] \label{cor: extreme}
    There exist a set of number $n_l \geq 0$, such that
    \begin{equation}
        p(N) =  \sum_l |n_l|, \quad N = \sum_l n_l F_l.
    \end{equation}
\end{corollary}
Besides, we can prove that the minimal value can be obtained on the decomposition into two points.
\begin{corollary}[two-point decomposition] \label{cor: pm}
    There exist two points $N_1, N_2 \in \mathcal{F}$, and $n_1, n_2 \geq 0$ such that 
    \begin{equation}
        p(N) = n_1 + n_2, \quad N = n_1 N_1 - n_2 N_2.
    \end{equation}
\end{corollary}\noindent
This corollary shows that the implementability function so defined is related to the robustness~\cite{RevModPhys.91.025001,PhysRevLett.115.070503} $R$ as
\begin{equation}
    p = 2 R + 1.
\end{equation}
\begin{proof}
    From Theorem~\ref{cor: extreme}, we have
    \begin{equation}
        N = \sum_l n_l F_l.
    \end{equation}
    Divide the extreme point as two set $\{F_l^+\}$ and $\{F_l^-\}$, where $n_l>0$ for $F_l^+$, and $n_l<0$ for $F_l^-$, and denote $n_l^{\pm} = |n_l|$ for $F_l^{\pm}$, we have
    \begin{equation}
        N = \sum_l n_l^+ F_l^+ - \sum_l n_l^- F_l^-.
    \end{equation}
    Let $n_1 = \sum_l n_l^+, n_2 = \sum_l n_l^-$, then
    \begin{equation}
        N_1 = \sum_l \frac{n_l^+}{n_1} F_l^+, N_2 = \sum_l \frac{n_l^-}{n_2} F_l^- \in \mathcal{F}
    \end{equation}
    for $\{F_l\}$ are extreme points.
\end{proof}

In the Lemma~\ref{lem: extreme}, we use the extreme point $\mathrm{ext}(\mathcal{F})$ to completely describe the convex free set $\mathcal{F}$, which is only possible for the bounded closed convex set.
Since the free set is not defined by sufficient properties like the case of physical implementability, we have to employ the extreme points that are intrinsic in the convex set, to describe the convex set, and assume the boundedness and closeness.
We will show that the set of extreme points with non-zero components is at most countable, so the summation is suitable.
\begin{proposition} \label{pro: countable}
    If the decomposition $N =  \sum_l n_l F_l$ is $L_1$ integrable, namely $\sum_l |n_l| < \infty$, then the set $A = \{F_l: n_l \neq 0\}$ of extreme points with non-zero component is at most countable.
\end{proposition}
\begin{proof}
    Let $A_m = \left\{F_l: |n_l| \geq \frac{1}{m}\right\}$, then the cardinal of the set $A_m$ is 
    \begin{equation}
        |A_m| = \sum_{A_m} 1 \leq m \sum_{A_m} |n_l| < \infty.
    \end{equation}
    Each $A_m$ is a finite set.
    Therefore, the set $A = \bigcup_{m} A_m$ is at most countable.
\end{proof}

\subsection{Properties}

Then, we consider some properties of $p(N)$.
Many of them are similar to the physical implementability.
\begin{proposition}[Faithfulness] \label{pro: faithfulness}
    \begin{equation}
        p(N) \geq 1, \quad N \in \mathcal{F} \Leftrightarrow p(N) = 1.
    \end{equation}
\end{proposition}
\begin{proof}
    With Corollary~\ref{cor: pm}, and $n_1 - n_2 = 1$, and $n_1, n_2 \geq 0$, we have 
    \begin{equation}
        p(N) = n_1 + n_2 = 1 + 2 n_2 \geq 1.
    \end{equation}
    If $N \in \mathcal{F}$, $p(N) = 1$ is obvious.
    On the contrary, if $p(N) = n_1 + n_2 = 1$, then $n_1 = 1, n_2 = 0$, so $N = N_1 \in \mathcal{F}$.
\end{proof} \noindent
The faithfulness here is a little different from the faithfulness of resource measure, but this has no harm.
\begin{proposition}[Sub-linearity] \label{pro: sub-linearity}
    \begin{equation}
        p(a N_1 + b N_2) \leq |a| p(N_1) + |b| p(N_2).
    \end{equation}
\end{proposition}
\begin{proof}
    Let the decomposition $N_1 = \sum_l n_{1l} F_l$ and $N_2 = \sum_l n_{2l} F_l$ reach $p(N_1)$ and $p(N_2)$, which is possible for the Theorem~\ref{the: min}.
    Then, the combination of $N_1$ and $N_2$ can be decomposed as
    \begin{equation}
        a N_1 + b N_2 = \sum_l (a n_{1l} + b n_{2l}) F_l,
    \end{equation}
    thus
    \begin{align}
        p(a N_1 + b N_2) & \leq \sum_l |a n_{1l} + b n_{2l}| \leq \sum_l |a||n_{1l}| + |b| |n_{2l}| \nonumber\\
        & = |a| p(N_1) + |b| p(N_2).
    \end{align}
\end{proof}

\begin{proposition}[Composition sub-multiplicity] \label{pro: sub-multiplicity}
    Let $N \in \mathcal{A}(A \rightarrow B), M \in \mathcal{A}(B \rightarrow C)$, then
    \begin{equation} \label{eq: sub-multiplicity}
        p(M \circ N) \leq p(M) p(N).
    \end{equation}
\end{proposition}
\begin{proof}
    Let the minimal decomposition of $N, M$ be
    \begin{align}
        N & = \sum n_i N_i, \\
        M & = \sum m_j M_j , 
    \end{align} 
    where $N_i \in \mathcal{F}(A \rightarrow B),  M_j \in \mathcal{F}(B \rightarrow C)$,
    then the composition is
    \begin{equation}
        M \circ N = \sum_{i,j} m_j n_i M_j \circ N_i.
    \end{equation}
    Since $\mathcal{F}$ is the free set, we have $M_j \circ N_i \in \mathcal{F}(A \rightarrow C)$, which means that $M \circ N$ can be decomposed into the elements of the free set $\mathcal{F}(A \rightarrow C)$.
    By the definition
    \begin{equation}
        p(M \circ N) \leq \sum_{i,j} |m_j n_i| = p(M) p(N).
    \end{equation}
\end{proof} \noindent
In particular, let $A = \mathbb{C}$, then $N = \rho \in \mathcal{B}(B)$, and $M \circ N = \mathcal{M}(\rho) \in \mathcal{B}(C)$.
\begin{corollary} \label{cor: state-operation bound}
    \begin{equation}
        \frac{p(\mathcal{M}(\rho))}{p(\rho)} \leq p(\mathcal{M}).
    \end{equation}
\end{corollary}

\begin{proposition}[Tensor-product sub-multiplicity] \label{pro: tensor-product}
    Let $M \in \mathcal{A}(X), N \in \mathcal{A}(Y)$.
    If the free set $\mathcal{F}(X), \mathcal{F}(Y)$ and $\mathcal{F}(XY)$ admits a tensor-product structure, then
    \begin{equation}
        p(M \otimes N) \leq p(M) p(N).
    \end{equation}
    In particular, if the free set $\mathcal{F}(XY) = \mathrm{conv}[\mathcal{F}(X) \otimes \mathcal{F}(Y)]$ is separable, the equality holds.
\end{proposition}\noindent
Here, $XY = A B$ if $X = A, Y = B$, and $XY = AB \rightarrow CD$ if $X = A \rightarrow C, Y = B \rightarrow D$.
\begin{proof}
    \textbf{1.} Sub-multiplicity: 

    Let the $p$ of $M, N$ is minimized by
    \begin{align}
        N & = \sum n_i N_i, \\
        M & = \sum m_j M_j , 
    \end{align}
    then
    \begin{equation}
        M \otimes N  = \sum_{i,j} m_j n_i M_j \otimes N_i.
    \end{equation}
    Since $\mathcal{F}$ admits a tensor-product structure, $M_j \otimes \mathrm{id}_Y, \mathrm{id}_X \otimes N_i \in \mathcal{F}(XY)$ are free, thus $M_j \otimes N_i \in \mathcal{F}(XY)$ are also free, and we have
    \begin{equation}
        p(M \otimes N) \leq p(M) p(N).
    \end{equation} 

    \noindent
    \textbf{2.} Equality if $\mathcal{F}(XY) = \mathrm{conv}[\mathcal{F}(X) \otimes \mathcal{F}(Y)]$: 

    Since $\mathcal{F}(XY) = \mathrm{conv}[\mathcal{F}(X) \otimes \mathcal{F}(Y)]$, the extreme points of $\mathcal{F}(XY)$ are $\{F_l^X \otimes F_k^Y\}$, where $\{F_l^X\}$ and $\{F_k^Y\}$ are extreme points of $\mathcal{F}(X)$ and $\mathcal{F}(Y)$.
    Let $p(M \otimes N)$ is attained at the decomposition
    \begin{align}
        M \otimes N & = \sum_{l,k} z_{lk} F_l^X \otimes F_k^Y \\
        & = \sum_l x_l F_l^X \otimes \left(\sum_k \frac{z_{lk}}{x_l} F_k^Y \right) \nonumber\\
        & = \sum_k \left(\sum_l \frac{z_{lk}}{y_k} F_l^X \right)\otimes y_k F_k^Y, \nonumber
    \end{align}
    where $x_l = \sum_k z_{lk}$ and $y_k = \sum_l z_{lk}$.
    Therefore, we have
    \begin{align}
        M = \sum_l x_l F_l^X = \sum_l \frac{z_{lk}}{y_k} F_l^X, \\
        N = \sum_k y_k F_k^Y = \sum_k \frac{z_{lk}}{x_l} F_k^Y.
    \end{align}
    By the definition of implementability function, we have
    \begin{align}
        p(M) & \leq \sum_l |x_l|, \quad 
        p(M)  \leq \sum_l \frac{|z_{lk_0}|}{|y_{k_0}|}, \\
        p(N) & \leq \sum_k |y_k|, \quad
        p(N)  \leq \sum_k \frac{|z_{l_0k}|}{|x_{l_0}|}.
    \end{align}
    For $p(M), p(N) \geq 0$, we have 
    \begin{align*}
        [p(M)p(N)]^2 & \leq \sum_{l_0, k_0} |x_{l_0}| |y_{k_0}| p(M)p(N) \\
        & \leq \sum_{l_0, k_0} |x_{l_0}| |y_{k_0}| \sum_l \frac{|z_{lk_0}|}{|y_{k_0}|} \sum_k \frac{|z_{l_0k}|}{|x_{l_0}|} \\
        & = \sum_{l, k_0} |z_{lk_0}| \sum_{l_0, k} |z_{l_0k}| = [p(M \otimes N)]^2.
    \end{align*}
    It means 
    \begin{equation}
        p(M)p(N) \leq p(M \otimes N),
    \end{equation}
    which closes the proof.
\end{proof}\noindent
We do not have $p(M \otimes N) = p(M) p(N)$ in general, which is held for physical implementability, because the free set of composition system $XY$ is not defined specifically.
However, with the constraint on the free set of composition system that it is separable, we can get equality.
We note that it is not broad enough, for example, the additivity of physical implementability $\nu$ is beyond this condition.
However, this condition is necessary, because the equality in the physical implementability is excluded by it. 

With these two sub-multiplicity, we also have the monotonicity under free superchannels 
\begin{equation}
    \Theta^{C\rightarrow D}(\mathcal{N}^{A\rightarrow B}) = \mathcal{P}^{BE \rightarrow D} \circ (N^{A \rightarrow B} \otimes \mathrm{id}^E) \circ \mathcal{Q}^{C \rightarrow AE},
\end{equation}
where both $\mathcal{P}^{BE \rightarrow D} \in \mathcal{F}(BE \rightarrow D), \mathcal{Q}^{C \rightarrow AE} \in \mathcal{F}(C \rightarrow AE)$ are free operations.
\begin{corollary}[Superchannel Monotonicity]
    \begin{equation}
        p\left(\Theta^{C\rightarrow D}(\mathcal{N}^{A\rightarrow B})\right) \leq p(\mathcal{N}^{A\rightarrow B}).
    \end{equation}
\end{corollary}

\begin{proposition} \label{pro: subset}
    If $\mathcal{F}_1 \subset \mathcal{F}_2$, then $p_{\mathcal{F}_1} \geq p_{\mathcal{F}_2}$.
\end{proposition}

\begin{proof}
    Let $N = \sum_{l} n_l F_l$ be the optimal decomposition with respect to extreme points of free set $\mathcal{F}_1$.
    Since $\mathcal{F}_1 \subset \mathcal{F}_2$, $F_l \in \mathcal{F}_2$, by the definition of implementability function
    \begin{equation}
        p_{\mathcal{F}_2}(N) \leq \sum_{l} |n_l| = p_{\mathcal{F}_1}(N).
    \end{equation}
\end{proof}

\subsection{Geometry property}

\begin{theorem} \label{the: geometry}
    The implementability function $p: \mathcal{A} \rightarrow \mathbb{R}$ is the Minkowski functional $p_{\mathcal{C}}: V \rightarrow \mathbb{R}$ of the convex closure $\mathcal{C} = \mathrm{conv}(\mathcal{F} \cup -\mathcal{F})$, where $-\mathcal{F} = \{- N: N \in \mathcal{F}\}$, when constrained on the affine space $\mathcal{A} = \mathrm{aff}(\mathcal{F})$
    \begin{equation}
        p = p_{\mathcal{C}}|_{\mathcal{A}}.
    \end{equation}
\end{theorem}

\begin{proof}
    Let the extreme points of $\mathcal{F}$ be $\mathrm{ext}(\mathcal{F}) = \{F_l\}$, then the extreme points of convex closure $\mathcal{C}$ is $\{F_l, -F_l\}$. Therefore, $0 \in \mathcal{C}$ and the Minkowski functional is so defined.

    For $N \in \mathcal{A}$, let the implementability function $p(N)$ is attained on the decomposition
    \begin{equation}
        N = \sum_l n_l F_l = p(N) \tilde{N},
    \end{equation}
    where $\tilde{N} = \sum_l q_l \mathrm{sgn}(n_l) F_l$ with the quasi-probability $q_l = \frac{|n_l|}{p(N)} \geq 0$. 
    For the quasi-probability is normalized, $\sum_l q_l = 1$, the element $\tilde{N}$ is the convex combination of $\{F_l, -F_l\}$, thus $\tilde{N} \in \mathcal{C}$, and $N \in p(N) \mathcal{C}$.
    By the definition of the Minkowski functional, we have
    \begin{equation}
        p_{\mathcal{C}}(N) \leq p(N).
    \end{equation}

    Conversely, since $\mathcal{F}$ is bounded, the convex closure $\mathcal{C}$ is also bounded, and with the same reasoning of Theorem~\ref{the: min}, we have that the infimum of $p_{\mathcal{C}}$ is attained.
    Let $p_{\mathcal{C}}(N)$ is attained on 
    \begin{equation}
        N = p_{\mathcal{C}}(N) \breve{N},
    \end{equation}
    then $\breve{N} \in \mathcal{C}$ can be decomposed as the convex combination of $\{F_l, -F_l\}$
    \begin{equation}
        \breve{N} = \sum_l \breve{q}_l (s_l F_l),
    \end{equation}
    where $\breve{q}_l \geq 0, \sum_l \breve{q}_l = 1$ and $s_l = \pm 1$.
    By the definition of the implementability function $p(N)$, we have
    \begin{equation}
        p(N) \leq \sum_l |p_{\mathcal{C}}(N) \breve{q}_l s_l| = p_{\mathcal{C}}(N).
    \end{equation}
\end{proof}

\begin{corollary} \label{cor: affine_invariant}
    The implementability function $p(N) = (N, \tilde{N},O) \equiv {|{ON}|}/{|O\tilde{N}|}$ is the ratio of $|{ON}|$ and $|O\tilde{N}|$, where $\tilde{N} = \partial \mathcal{C} \cap ON$ is the intersection of the boundary $\partial \mathcal{C}$ with the line $ON$.
    Moreover, it is the affine invariant in the unnormalized space $V$, and the projective invariant of normalized space $\mathcal{A}$.  
\end{corollary}
We can thus give a bound of the implementability function
\begin{corollary}
    \begin{equation}
        \frac{\Vert N \Vert_2}{r_{\max}} \leq p(N) \leq \frac{\Vert N \Vert_2}{r_{\min}},
    \end{equation}
    where $r_{\max} = \max \Vert F_l \Vert_2$, and $r_{\min} = \inf \{r: B(0, r) \subset \mathcal{C}\}$.
\end{corollary}  
The theorem~\ref{the: geometry} says that the Minkowski functional $p_{\mathcal{C}}$ is the linear scalable extension of the implementability function $p(N)$ to the vector space $V$ by $p(a N) = |a| p(N)$.
Therefore, we need not distinguish the implementability function $p_{\mathcal{F}}$ and the Minkowski functional $p_{\mathcal{C}}$ in the following, where $\mathcal{C} = \mathrm{conv}(\mathcal{F} \cup -\mathcal{F})$.
In this sense, the convex set $\mathcal{C}$ can be interpreted as the extended free set in the unnormalized resource space.

\begin{proposition}
    The implementability function $p_{\mathcal{C}}$ is the base norm $\Vert\cdot \Vert_{\mathcal{F}}$ of the vector space $V$.
\end{proposition}
\begin{proof}
    The sub-linearity is shown in the Proposition~\ref{pro: sub-linearity}, and the non-negative is from the definition.
    Assume $p_{\mathcal{C}}(\tilde{\mathcal{N}}) = 0$ and $\tilde{\mathcal{N}} \neq 0$, then there is a neighborhood $B(0,r)$ separate $\tilde{\mathcal{N}}$ and $0$, namely $\Vert\tilde{\mathcal{N}}\Vert_2 > r$.
    Then We consider the operation $a \tilde{\mathcal{N}}$, the implementability function
    \begin{equation}
        p_{\mathcal{C}}(a \tilde{\mathcal{N}}) \geq \frac{|a| \cdot \Vert\tilde{\mathcal{N}}\Vert_2}{r_{\max}} > \frac{|a| r}{r_{\max}} >0,
    \end{equation} 
    since $\mathcal{C}$ is bounded, and $r_{\max} < \infty$.
    On the contrary, the scalability shows that $p_{\mathcal{C}}(a \tilde{\mathcal{N}}) = a p_{\mathcal{C}}(\tilde{\mathcal{N}}) = 0$, the contradiction proves that $\tilde{\mathcal{N}} = 0$.
    Therefore, $p_{\mathcal{C}}$ is a norm.
    By Theorem~\ref{the: geometry}, the norm $p_{\mathcal{C}} = \Vert\cdot \Vert_{\mathcal{F}}$ is the base norm.
\end{proof}

\begin{corollary}
    The algebra $V(A \rightarrow A)$ with operation composition is a Banach algebra with respect to the base norm $\Vert\cdot \Vert_{\mathcal{F}}$.
\end{corollary}
\begin{proof}
    Proposition~\ref{pro: sub-multiplicity} shows that $V(A \rightarrow A)$ is a Banach algebra. 
\end{proof}

\section{Optimal decompositions} \label{app: decomposition}

    In the proof of the Theorem~\ref{the: min}, there is an ambient structure of $\mathcal{A}$, or the normalized space $\mathcal{A} = V/\mathbb{R}$.
    Let $\{G_l\}$ be a set of affine independent points, whose cardinal is the same as $\mathrm{ext}(\mathcal{F}) = \{F_l\}$.
    Let $\phi: F_l \mapsto G_l$ is the bijective between $\{F_l\}$ and $\{G_l\}$.
    Since $\{F_l\}$ are not always affine independent, there may be additional linear constraints $\{C_s = 0\}$, so we have
    \begin{equation}
        \mathcal{A} \simeq \mathrm{aff}(\{G_l\})/\langle\{\phi(C_s)\}\rangle.
    \end{equation}
    On the vector space $\langle\{G_l\}\rangle$, there is $L_1$ norm $\Vert \cdot \Vert_1$.
    The implementability function $p(N) = \Vert N + \langle\{\phi(C_s)\}\rangle \Vert_1$ is the $L_1$ distance between the plane $N + \langle\{\phi(C_s)\}\rangle$ and origin.
    Actually, the projection $\pi: \mathrm{aff}(\{G_l\}) \rightarrow \mathcal{A}$ defines a vector bundle, where $\mathcal{A}$ is the base manifold, and $\pi^{-1}(N) \simeq \langle\{\phi(C_s)\}\rangle$ is the fiber. 
    The coordinate transformations of a base manifold $\mathcal{A}$, namely the unitary dynamical resource $U \in \mathcal{F}(A \rightarrow A)$ for static resource $X = A$ and for dynamical resource $X = A \rightarrow B$ or $X = B \rightarrow A$, induce the translation function of vector bundle by its tautological representation.
    In this vector bundle, any point $(N, (x_s)) \in \mathrm{aff}(\{G_l\}) \simeq \mathcal{A} \times \langle\{\phi(C_s)\}\rangle$ uniquely represent a decomposition of the resource $N$.
    Therefore, $\mathcal{A} \times \langle\{\phi(C_s)\}\rangle$ is the space of decomposition.      
    
    \subsection{Uniqueness of optimality}
    In the space of decomposition, the optimal decompositions are of most interest.   
    The existence of optimal decomposition is proved in the Theorem~\ref{the: min}. 
    Then, it is of most interest that in what sense the optimal decomposition has the uniqueness.

    For $N \in \mathcal{A}$, let the different optimal decompositions with extreme points of $\mathcal{F}$, labeled by index $r$, be
    \begin{equation}
        N = \sum_l n_l^r F_l.
    \end{equation}
    Each of the optimal decompositions, determine a partition $(A_r^+, A_r^-, A_r^0)$ of the extreme points
    \begin{align}
        A_r^+ = \{F_l: n_l^r > 0\}, \\
        A_r^- = \{F_l: n_l^r < 0\}, \\
        A_r^0 = \{F_l: n_l^r = 0\}.
    \end{align} 
    There is a partial order on these partitions that $(A^+, A^-, A^0) \prec (A^{'+}, A^{'-}, A^{'0})$ if $A^+ \subset A^{'+}$ and $A^- \subset A^{'-}$.
    \begin{definition} \label{def: partition}
        The maximality over all the partitions $(A_r^+, A_r^-, A_r^0)$ of point $N$ is $(\breve{A}^+, \breve{A}^-, \breve{A}^0)$, where 
        \begin{align}
            \breve{A}^+ &= \bigcup_r A^+_r, \\
            \breve{A}^- &= \bigcup_r A^-_r, \\
            \breve{A}^0 &= A/(\breve{A}^+ \cup \breve{A}^-).
        \end{align}     
    \end{definition}
    \begin{lemma} \label{lem: separation}
        Let $N = \sum_l x_l F_l$ and $N = \sum_l y_l F_l$ be two different optimal decompositions, then $x_l y_l \geq 0$ for all $l$.
    \end{lemma} 
    \begin{proof}
        We first show that $x_l y_l < 0$ is impossible.
        We divide the extreme points into partition 
        \begin{align}
            C_{+} = \{F_l: x_l y_l \geq 0\},\\
            C_{-} = \{F_l: x_l y_l < 0\}.
        \end{align}
        Consider the convex combination of the two decompositions $\sum_l (a x_l + b y_l) F_l$, 
        \begin{align}
            \sum_l |a x_l + b y_l| = \sum_{C_+} (a |x_l| + b |y_l|) + \sum_{C_-} |a x_l + b y_l| \nonumber \\
            = p(N) - \sum_{C_-} (a |x_l| + b |y_l|) + \sum_{C_-} |a x_l + b y_l|.
        \end{align} 
        This is because for $l \in C_+$, $x_l$ and $y_l$ have the same signatures.
        Then, we divide the set $C_-$ into two sets,
        \begin{align}
            M^+ = \left\{F_l: \frac{|y_l|}{|x_l|} > \frac{a}{b}\right\}, \\
            M^- = \left\{F_l: \frac{|y_l|}{|x_l|} < \frac{a}{b}\right\}.
        \end{align}
        In the set $M^+$, $a x_l + b y_l$ and $x_l$ have different signatures, while having the same signature in the set $M_-$, thus 
        \begin{align}
            \sum_{M^+} |a x_l + b y_l| = \sum_{M^+} (b |y_l| - a |x_l|), \\
            \sum_{M^-} |a x_l + b y_l| = \sum_{M^+} (a |x_l| - b |y_l|).
        \end{align}
        Therefore, we have
        \begin{equation}
            \sum_l |a x_l + b y_l| = p(N) -  2 \left(a \sum_{M^+} |x_l| + b \sum_{M^-} |y_l| \right), 
        \end{equation} 
        which is less than $p(N)$ if $C_-$ is not empty.
    \end{proof}\noindent
    This result means that for arbitrary two optimal decompositions, labeled as $r$ and $r'$, the intersections $A_r^{\pm} \cap A_{r'}^{\mp} = 0$ are empty.

    In addition, the convex set generated by $A^+$ and $A^-$ are exclusive (the label of different decompositions $r$ is neglected if not necessary). 
    \begin{lemma} \label{lem: exclusive}
        If $N = \sum_l n_l F_l$ is the optimal decomposition, then $\mathrm{conv} A^+ \cap \mathrm{conv} A^- = \emptyset$.
    \end{lemma}
    \begin{proof}
        Assume the decomposition attains the minimum with $\mathrm{conv} A^+ \cap \mathrm{conv} A^- \neq \emptyset$, then there exist $R \in \mathrm{conv} A^+ \cap \mathrm{conv} A^-$, and it can be decomposed as $R = \sum_{A^+} a_k F_k = \sum_{A^-} b_l F_l$, where $a_k, b_l \geq 0, \sum_k a_k = \sum_l b_l = 1$.
        Consider the decomposition 
        \begin{equation}
            N = \sum_{A^+} (n_l - x a_l) F_l  - \sum_{A^-} (|n_l| - x b_l) F_l.
        \end{equation}
        If the set $A = A^+ \cup A^-$ is finite, then all $|n_l| \geq \underline{n}  = \min |n_l| > 0$, denote $C = \max \{a_l, b_k\}$, then let $0< x < \underline{n}/C$, we have
        \begin{align}
            & \sum_{A^+} |n_l - x a_l| + \sum_{A^-} ||n_l| - x b_l| \nonumber \\
            & = \sum_l |n_l| - 2 x < \sum_l |n_l|,
        \end{align}
        which contradicts the assumption of minimum.

        Otherwise, if the set $A$ is infinite, from the proof of Proposition~\ref{pro: countable}, every set $A_m$ where $|n_l| \leq \frac{1}{m}$ is finite, then the infimum $\inf |n_l| = 0$.
        In this case, we decompose the set $A^+,A^-$ by the partition $B^+_m = \left\{F_l: \frac{1}{m} \leq n_l < \frac{1}{m-1}\right\}$ and $B^-_m = \left\{F_l: \frac{1}{m} \leq -n_l < \frac{1}{m-1}\right\}$.
        In this partition, the series
        \begin{align}
            \sum_m \alpha_m = \sum_k a_k = 1, \\
            \sum_m \beta_m  = \sum_k b_k = 1,
        \end{align}
        where $\alpha_m = \sum_{B^+_m} a_k, \beta = \sum_{B^-_m} b_k$ convergent, thus $\lim_{M\rightarrow \infty} \sum_{m>M} \alpha_m = \lim_{M\rightarrow \infty} \sum_{m>M} \beta_m = 0$.
        Given $M$, let $x = \frac{1}{C N}$, then
        \begin{align} 
            & \sum_{A^+} |n_l - x a_l| + \sum_{A^-} ||n_l| - x b_l| \nonumber\\
            & = \left(\sum_{m<N} - \sum_{m>N}\right) \left[\sum_{B^+_m} (n_l - x a_l) + \sum_{B^-_m} (|n_l| - x b_l)\right] \nonumber\\
            & = \left(\sum_{m<N} - \sum_{m>N}\right) \left[\sum_{B_m} |n_l| - x (\alpha_m + \beta_m)\right] \nonumber\\
            & = \sum_l |n_l| - 2 \sum_{m>N} \sum_{B_m} |n_l| - 2 x \left[1 - \sum_{m>N} (\alpha_m + \beta_m)\right].
        \end{align}
        Since $\lim_{M\rightarrow \infty} \sum_{m>M} \alpha_m = \lim_{M\rightarrow \infty} \sum_{m>M} \beta_m = 0$, for any $\epsilon > 0$, there exists $M$, if $N \geq M$, then $\sum_{m>N} \alpha_m \leq \epsilon, \sum_{m>N} \beta_m \leq \epsilon$. 
        Therefore, 
        \begin{align} 
            & \sum_{A^+} |n_l - x a_l| + \sum_{A^-} ||n_l| - x b_l| \nonumber\\
            & \leq \sum_l |n_l| - 2 \sum_{m>N} \sum_{B_m} |n_l| - 2 x (1 - 2 \epsilon) \nonumber\\
            & < \sum_l |n_l| - 2 x (1 - 2 \epsilon),
        \end{align}
        which contradict with the assumption of minimum when $\epsilon \leq \frac{1}{2}$.
    \end{proof}

    \begin{theorem} \label{the: uniqueness}
        The maximality $(\breve{A}^+, \breve{A}^-, \breve{A}^0)$ of the partitions $(A_r^+, A_r^-, A_r^0)$ is also a partition of extreme points, and the generated convex sets $\mathrm{conv} \breve{A}^\pm$ are exclusive
        \begin{equation}
            \mathrm{conv} \breve{A}^+ \cap \mathrm{conv} \breve{A}^+ = \emptyset.
        \end{equation}
    \end{theorem}
    \begin{proof}
        Lemma~\ref{lem: separation} shows that $(\breve{A}^+, \breve{A}^-, \breve{A}^0)$ also form a partition.
        The exclusive property of convex sets $\mathrm{conv}A^\pm_r$, proved in the Lemma~\ref{lem: exclusive}, is inherited by $\mathrm{conv} \breve{A}^\pm$, with the fact that the convex combinations of optimal decompositions are also optimal decompositions.
        Give any sets $A^{\prime+} \subset \mathrm{conv} \breve{A}^+, A^{\prime-} \subset \mathrm{conv} \breve{A}^-$, there exists a sequence of partition $(A^+_{r_i}, A^-_{r_i}, A^0_{r_i})$, that $A^{\prime+} \subset \bigcup_i A^+_{r_i}$ and $A^{\prime-} \subset \bigcup_i A^-_{r_i}$. 
        The convex combination of the optimal decompositions labeled as $r_i$, whose coefficients of these optimal decompositions are non-zero, are also optimal decompositions, and the corresponding partition of extreme points is just $\left(\bigcup_i A^+_{r_i}, \bigcup_i A^-_{r_i}, \bigcap_i A^0_{r_i}\right)$.
        By the Lemma~\ref{lem: exclusive}, the convex closures of $\bigcup_i A^+_{r_i}$ and $\bigcup_i A^-_{r_i}$ are exclusive, so do $A^{\prime+}$ and $A^{\prime-}$.
        If $\mathrm{conv} \breve{A}^+ \cap \mathrm{conv} \breve{A}^+ \neq \emptyset$, then there exist $A^{\prime+}$ and $A^{\prime-}$ have intersection. 
    \end{proof} \noindent
    Therefore, for the unique partition of extreme points $(\breve{A}^+, \breve{A}^-, \breve{A}^0)$, the generated convex sets $\mathrm{conv}\breve{A}^+$ and $\mathrm{conv} \breve{A}^-$ can be separated by linear functional (family of hyperplanes), by separation theorem~\cite{prasolov2001geometry,hug2020lectures}.

    \subsection{Separation by hyperplanes}
    
    It is a problem what the linear functionals separating the two sets $\breve{A}^+$ and $\breve{A}^-$ are.
    To discuss this problem, we should consider the geometry properties of the convex sets $\mathrm{conv} \breve{A}^+$ and $\mathrm{conv} \breve{A}^-$.
    We begin with a two-point optimal decomposition.
    \begin{lemma} \label{lem: internal}
        For $N \in \mathcal{A} \setminus \mathcal{F}$, if the decomposition $N = a N^+ - b N^-$ is optimal, then $N^+, N^- \in \partial \mathcal{F}$.
    \end{lemma}
    \begin{proof}
        The two point $N^+,N^-$ determine a line $l$, which intersects the convex free set $\mathcal{F}$, since $N^+,N^- \in \mathcal{F}$.
        The intersection $l \cap \mathcal{F}$ is also a convex set, and thus a line segment.
        Let the endpoints of this segment are $N_1,N_2$, what we want to prove is $\{N^+, N^-\} = \{N_1,N_2\}$.
        Let $N^+ = x N_1 + (1- x) N_2, N^- = y N_1 + (1-y) N_2$, where $0 \leq y < x\leq1$.
        \begin{equation}
            N = (a x - b y) N_1 + [1 - (a x - b y)] N_2,
        \end{equation}
        where 
        \begin{align}
            & |a x - b y| + |1 - (a x - b y)| \nonumber \\ 
            & = \max\{2(a x - b y) - 1, 1\} \leq 2(a x - b y) - 1 \nonumber \\
            & \leq 2 a - 1 = a+ b = p(N),
        \end{align}
        where the equality is attained when $x = 1, y=0$, namely $N^+ = N_1, N^- = N_2$.
    \end{proof}
    \begin{theorem}\label{the: para}
        For $N \in \mathcal{A} \setminus \mathcal{F}$, the convex sets $\mathrm{conv} \breve{A}^{\pm} \subset \mathcal{P}^{\pm}$ are contained in two supporting hyperplanes $\mathcal{P}^{\pm}$ of the free set $\mathcal{F}$.
        In particular, if there is a two-point optimal decomposition $(N^+, N^-)$ is attained on smooth point $N^{\pm}$ of boundary, then the two hyperplanes are parallel $\mathcal{P}^{+} \parallel \mathcal{P}^{-}$.            
    \end{theorem}
    \begin{proof}
        \textbf{1.} Supporting: 

        Since the convex combinations of the optimal decompositions are also optimal decompositions, there is an optimal decomposition that has positive coefficients on all extreme points in $\breve{A}^+$ and negativity coefficients on $\breve{A}^-$.
        Denote this decomposition as 
        \begin{equation}
            N = \sum_{\breve{A}^+} n^+_l F_l - \sum_{\breve{A}^-} n^-_l F_l = n^+ N^+ - n^- N^-,
        \end{equation}
        where $N^{\pm} = \sum_{\breve{A}^{\pm}} \frac{n^{\pm}_l}{n^{\pm}} F_l, n^{\pm} = \sum_{\breve{A}^{\pm}} n^{\pm}_l$.
        It is clear that $N^{\pm} \in \mathrm{relint}(\mathrm{conv} \breve{A}^{\pm})$ is in the relative interior of the convex set $\mathrm{conv} \breve{A}^{\pm}$.
        If the extreme points $\breve{A}^{\pm}$ can span an affine space with dimension $n = \dim \mathcal{A}$, then there is a neighborhood $D^{\pm} \subset \mathrm{int}(\mathrm{conv} \breve{A}^{\pm}) \subset \mathcal{F}$ of $N^{\pm}$.
        While by the Lemma~\ref{lem: internal}, we have $N^{\pm} \in \partial \mathcal{F}$, which means no neighborhood of $N^{\pm}$ is subset of $\mathcal{F}$.
        This contradiction proves that $\breve{A}^{\pm}$ are contained in hyperplanes of $\mathcal{A}$.

        Let the two hyperplanes be $\mathcal{P}^{\pm}$, assume they are not supporting hyperplanes~\cite{prasolov2001geometry,hug2020lectures}, there is a point $M^{\pm} \in \mathcal{P}^{\pm} \cap \mathcal{F}$ that $M \not \in \partial \mathcal{F}$.
        So, there is a neighborhood $D(M^{\pm}, \epsilon) \subset \mathcal{F}$ of $M^{\pm}$.
        The line $l^{\pm}$ through $N^{\pm}$ and $M^{\pm}$ have the intersection $l \cap \mathcal{F}$ with $\mathcal{F}$, which is also a convex set, i.e. line segment.
        Let the endpoint $P^{\pm} \in \mathcal{F}$ be the one closer to $N^{\pm}$ than $M^{\pm}$.
        The convex closure $\mathrm{conv}(D(M^{\pm}, \epsilon) \cup \{P^{\pm}\}) \subset \mathcal{F}$ for the convexity.
        Then, all the neighborhood $D(N^{\pm}, \epsilon')$ of $N^{\pm}$ where the radius $\epsilon' \leq \frac{|N^{\pm}P^{\pm}|}{|M^{\pm}P^{\pm}|} \epsilon$ are subsets of $\mathcal{F}$, which contradicts with $N^{\pm} \in \partial \mathcal{F}$.
        
        \noindent
        \textbf{2.} Parallel: 

        It is clear that for any two-point decomposition of $N$, $N^{\pm}$ and $N$ are on a straight line.
        Therefore, we start from an arbitrary line $l$ though the point $N$ intersects the boundary $\partial \mathcal{F}$.
        We assume the intersection is two different points, otherwise it cannot give a decomposition of $N$.
        Denote the intersecting point closer to $N$ as $N^-$, another as $N^+$, then we define a pre-implementability $p = 2 b + 1$, where $b = |N^- N|/|N^- N^+|$.    
        This 

        If both $N^{\pm}$ are smooth, we consider the tangent hyperplanes $\mathcal{T}_{N^{\pm}}$ of $N^{\pm}$, which are the unique supporting hyperplanes at these two points.
        Assuming these two hyperplanes $\mathcal{T}_{N^{\pm}}$ are not parallel, then they have an intersection which is a plane of dimension $n-2$.
        Consider the quotient space $\mathcal{A}/(\mathcal{T}_{N^{-}} \cap \mathcal{T}_{N^{-}})$, namely the $2$ dimensional sector of $\mathcal{A}$ which contains the line $l$ and a point $M \in \mathcal{T}_{N^{-}} \cap \mathcal{T}_{N^{-}}$.
        Denote the acute angles between $l$ and $MN^{\pm}$ as $0 < \alpha^{\pm} \leq \pi/2$.
        Give a continuous parameter $t$, we translate the point $N^{+}$ on the boundary along the direction of $\overrightarrow{M N^{+}}$.
        Then, the line $N N^{+}_t$, denoted as $l_t$, intersects the boundary at another point $N^{-}_t$.
        They also form a decomposition of $N$ with $p(t) = 2 b(t) + 1$, where $b(t) = |N^-_t N|/|N^-_t N^+_t|$.
        Denote the projection of $N^{\pm}$ on $l_t$ at $\tilde{N}^{\pm}$.
        The diagram is shown in Supplementary Fig.~\ref{fig: plane_geo}.

        Reparameterizing $t$ so that $|N^+_t N^+| = t + O(t^2)$, then $$|N^+ \tilde{N}^+| = t \sin \alpha^+ + O(t^2), \\ |N^+_t \tilde{N}^+| = t \cos \alpha^+ + O(t^2).$$
        Since $N^{\pm} \tilde{N}^{\pm} \perp l_t$, they are parallel $N^{+} \tilde{N}^{+} \parallel N^{\pm} \tilde{N}^{\pm}$.
        By the definition of $b$, we have $|N^- \tilde{N}^-| = t \frac{b}{1 + b} \sin \alpha^+ + O(t^2)$, thus $|N^-_t \tilde{N}^-| = t \frac{b}{1 + b} \frac{\sin \alpha^+}{\tan \alpha^-} + O(t^2)$.
        Besides, for convenience, denote $|N^+ N^-| = 1$, then $|\tilde{N}^+ \tilde{N}^-| = 1 + O(t^2), |N \tilde{N}^-| = b + O(t^2)$.
        By the definition of $b(t)$ and $p(t)$, we have
        \begin{align}
            b(t) & = \frac{b - t \frac{b}{1 + b} \frac{\sin \alpha^+}{\tan \alpha^-} + O(t^2)}{1 + t \left(\cos \alpha^+ + \frac{b}{1 + b} \frac{\sin \alpha^+}{\tan \alpha^-}\right) + O(t^2)}, \nonumber \\
            & = b \left[1 - t \left(\cos \alpha^+ + \frac{\sin \alpha^+}{\tan \alpha^-}\right)\right] + O(t^2), \\
            \dot{p} & = - (p - 1)\left(\cos \alpha^+ + \frac{\sin \alpha^+}{\tan \alpha^-}\right) \leq 0.
        \end{align}
        We optimize the $N^{\pm}$ continuously until attaining an optimal decomposition.
        If this optimal decomposition is attained at a smooth point, then $\dot{p} = 0$, either $p = 1$ or $\alpha^+ + \alpha^- = \pi$.
        The former means the trivial case where $N \in \mathcal{F}$ and there is no $N^-$ point, which is excluded from our consideration, while the latter means $\alpha^{\pm} = \pi/2$, namely the two hyperplanes are parallel $\mathcal{T}_{N^{+}} \parallel \mathcal{T}_{N^{+}}$. 
        Note that for optimal decomposition $\mathcal{T}_{N^{\pm}} = \mathcal{P}^{\pm}$, the contradiction proves what we want.
    \end{proof} 
    \begin{figure}[t]
        \centering
        \includegraphics[width=0.48\textwidth]{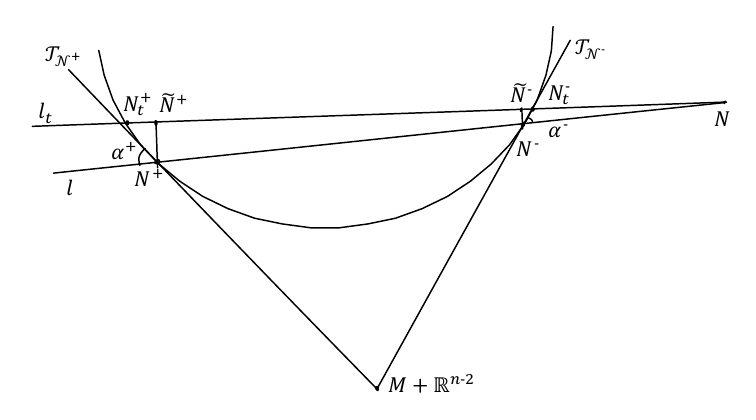}
        \caption{\textbf{The diagram of Theorem~\ref{the: para}.}}
        \label{fig: plane_geo}
    \end{figure}
    In the above, we find that the two convex sets $\mathrm{conv} \breve{A}^{\pm}$ are subsets of supporting sets of the convex free set $\mathcal{F}$.
    Therefore, there are two linear functions $\mathcal{P}^{\pm}$ take the same value on the convex sets $\mathrm{conv} \breve{A}^{\pm}$ correspondingly, and in certain cases, these two linear functional are same.
    This result shows that the necessary condition to attain optimal two-point decomposition is the supporting hyperplanes of the points are parallel or one of the two points located on vertices of the set $\mathcal{F}$.
    Therefore, if there is an optimal two-point decomposition not located on the vertices of the free set $\mathcal{F}$, the separating linear functional is dual to the normal vector of the supporting hyperplanes of the two points of decomposition.
    It allows us to check whether one two-point decomposition is optimal.
    Moreover, it also gives a lower bound of the implementability.
    \begin{corollary}
        Let $m(N, f) = \min \{f(E - N): E \in \mathcal{F}\}$ and $M(N, f) = \max \{f(E - N): E \in \mathcal{F}\}$, where $f: \mathcal{A} \rightarrow \mathbb{R}$ denotes linear functional on $\mathcal{A}$, then
        \begin{equation}
            p(N) \geq \min_{f} \frac{m(N, f)}{M(N, f) - m(N, f)}.
        \end{equation}
    \end{corollary}

    \subsection{General solution of optimal decomposition}

    Then, we consider how to construct all the optimal decompositions.
    As the proof of corollary~\ref{cor: pm}, any decomposition with respect to extreme points uniquely corresponds to a two-point decomposition $N = n^+ N^+ - n^- N^-$, and conversely, with the convex decomposition of $N^{\pm}$ into extreme points, the optimal extreme-point decomposition can be constructed from the optimal two-point decomposition.
    Therefore, all the optimal two-point decompositions are sufficient.
    We are interested in the set of all optimal two-point decomposition $(N^+, N^-)$, denoted as $D$.
    Obviously, $D \subset \mathrm{conv} \breve{A}^{+} \times \mathrm{conv} \breve{A}^{-}$.
    We construct it in detail. 
    \begin{theorem} \label{the: general_solution}
        Let $N = n^{+} N^+ - n^- N^-$ be an optimal decomposition.
        Denoting
        \begin{equation}
            Q = n^+(\mathrm{conv} \breve{A}^{+} - N^+) \cap n^-(\mathrm{conv} \breve{A}^{-} - N^-),
        \end{equation}
        the set of all optimal two-point decomposition 
        \begin{equation}
            D = \{(N^+ + M / n^+, N^- + M / n^-): M \in Q\}.
        \end{equation}            
    \end{theorem}
    \begin{proof}
        Let $(\tilde{N}^+, \tilde{N}^-) \in D$, then $n^{+} (\tilde{N}^+ - N^+ ) = n^{-} (\tilde{N}^- - N^-) \equiv M$. 
        Since $\tilde{N}^{\pm} \in \mathrm{conv} \breve{A}^{\pm}$, $M \in Q$, which means that
        \begin{equation}
            D \subset \{(N^+ + M / n^+, N^- + M / n^-): M \in Q\}.
        \end{equation}
        In the contrary, let $M \in Q$, then $N^{\pm} + M / n^{\pm} \in \mathrm{conv} \breve{A}^{\pm} \subset \mathcal{F}$, and
        \begin{equation}
            n^{+} (N^+ + M / n^+) - n^- (N^- + M / n^-) = N
        \end{equation} 
        is an optimal decomposition.
        Therefore, $(N^+ + M / n^+, N^- + M / n^-) \in D$.
    \end{proof}
    It is possible in general that $\mathrm{conv} \breve{A}^{\pm} \subsetneq \mathcal{P}^{\pm} \cap \partial \mathcal{F}$ cannot be a supporting set.
    However, in this theorem, the sets $\mathrm{conv} \breve{A}^{\pm}$ can be harmlessly substituted with $\mathcal{P}^{\pm} \cap \partial \mathcal{F}$.
    \begin{proposition} \label{pro: supporting}
        $\mathrm{conv} \breve{A}^{\pm} \subset \mathcal{P}^{\pm} \cap \partial \mathcal{F}$. 
        Moreover, if $\dim \mathrm{conv} \breve{A}^{\pm} = \dim \mathcal{P}^{\pm} \cap \partial \mathcal{F}$, then $\mathrm{conv} \breve{A}^{\pm} = \mathcal{P}^{\pm} \cap \partial \mathcal{F}$.
    \end{proposition}
    \begin{proof}
        The inclusion is obvious.
        Assume $\mathrm{conv} \breve{A}^{\pm} \subsetneq \mathcal{P}^{\pm} \cap \partial \mathcal{F}$ and $\dim \mathrm{conv} \breve{A}^{\pm} = \dim \mathcal{P}^{\pm} \cap \partial \mathcal{F}$.
        There are $d^{\pm} + 1$ affine independent extreme points $\{F_{l_k}^{\pm}, k = 1, 2, \dots d^{\pm} + 1\} \subset \breve{A}^{\pm}$, and there is extreme point $F_{l_0}^{\pm} \in \mathcal{P}^{\pm} \cap \partial \mathcal{F} \setminus \mathrm{conv} \breve{A}^{\pm}$.
        Let $N^{\pm} \in \mathrm{conv} \breve{A}^{\pm}$ attain optimal decomposition where there is a convex decomposition of $N^{\pm}$ is nonzero on $\{F_{l_k}^{\pm}, k = 1, 2, \dots d^{\pm} + 1\} \subset \breve{A}^{\pm}$
        \begin{equation}
            N^{\pm} = \sum_{k} n_k^{\pm} F_{l_k}^{\pm} + \sum_{\breve{A}^{\pm} \setminus \{F_{l_k}^{\pm}\}} n_l^{\pm} F_l,
        \end{equation}
        with $n_k > 0$.
        The extreme point $F_{l_0}^{\pm}$ can be affine decomposed into the independent extreme points $\{F_{l_k}^{\pm}, k = 1, 2, \dots d^{\pm} + 1\} \subset \breve{A}^{\pm}$ as
        \begin{equation}
            C_0^{\pm} = F_{l_0}^{\pm} - \sum_{k} f_k^{\pm} F_{l_k}^{\pm} = 0.
        \end{equation}
        Let $0 < x^{\pm} \leq \frac{\min n_k^{\pm}}{\max f_k^{\pm}}$, then 
        \begin{equation}
            N^{\pm} = \sum_{k} n_k^{\pm} F_{l_k}^{\pm} + \sum_{\breve{A}^{\pm} \setminus \{F_{l_k}^{\pm}\}} n_l^{\pm} F_l + x C_0^{\pm}
        \end{equation}
        is a convex decomposition of $N^{\pm}$ where the coefficient $x$ of $F_{l_0}^{\pm}$ is larger than zero, which contradicts the definition of $\breve{A}^{\pm}$.
    \end{proof}\noindent
    Combining with proposition~\ref{pro: supporting}, we have the following corollary.
    \begin{corollary}
        The quotient space $\langle\mathcal{P}^{\pm} \cap \partial \mathcal{F}\rangle/\langle\mathrm{conv} \breve{A}^{\pm}\rangle$ are the subspace of the complement space of $\langle Q\rangle$
        \begin{equation}
            \langle\mathcal{P}^{\pm} \cap \partial \mathcal{F}\rangle/\langle\mathrm{conv} \breve{A}^{\pm}\rangle \subset V/\langle Q\rangle.
        \end{equation}
    \end{corollary}

    \begin{corollary}
        \begin{equation}
            \dim D = \dim Q \leq \min\{\dim \breve{A}^{+},\dim \breve{A}^{-}\}.
        \end{equation}
    \end{corollary}

    \begin{corollary}
        If one optimal decomposition $N = n^+ N^+ - n^- N^-$ is attained where one of $N^{\pm}$ is an extreme point which is not the endpoint of any line segment in boundary $\partial \mathcal{F}$, this optimal two-point decomposition is unique.
        If both $N^{\pm}$ are extreme points that are not the endpoints of any line segment in boundary $\partial \mathcal{F}$, this optimal extreme-point decomposition is unique.
        In particular, if the free set $\mathcal{F}$ is strictly convex, then the optimal decomposition with respect to extreme points is unique.
    \end{corollary}
    \begin{proof}
        By the assumption, that one of $N^{\pm}$ is an extreme point, then this point locates on one of the supporting planes $\mathcal{P}^{\pm}$.
        However, this extreme point is not the endpoint of any line segment in boundary $\partial \mathcal{F}$, thus the corresponding supporting plane intersects the boundary $\partial \mathcal{F}$ only at this point, which means that $\min\{\dim \breve{A}^{+},\dim \breve{A}^{-}\} = 0$.
        Thus, $\dim D = 0$, which means the uniqueness of the optimal two-point decomposition. 

        Moreover, if both $N^{\pm}$ is an extreme point which is not the endpoint of any line segment in boundary $\partial \mathcal{F}$, then both $\dim \breve{A}^{\pm} = 0$, and both $\mathrm{conv} \breve{A}^{\pm}$ contain only a single extreme point.
        Therefore, this decomposition is the only optimal extreme-point decomposition.
        In particular, if the free set $\mathcal{F}$ is strictly convex, then every extreme point is not the endpoint of any line segment in boundary $\partial \mathcal{F}$, and the set of extreme points $\mathrm{ext} \mathcal{F} = \partial \mathcal{F}$.
        Therefore,  both $N^{\pm}$ is an extreme point that is not the endpoint of any line segment in boundary $\partial \mathcal{F}$.
    \end{proof}

    Theorem~\ref{the: general_solution} gives the way to construct all the optimal decomposition from a specific optimal decomposition with the help of the two hyperplanes $\mathcal{P}^{\pm}$.
    With more calculations, we find that these two hyperplanes should be parallel in many cases.
    Note that for $N$, the pre-implementability $p = g[(x_s)]$ of the two-point decomposition $N = n^+ N^+ - n^- N^-$ is a function of $(N^+, N^-) \in \mathcal{F}^2$, and since the optimal decompositions are attained only on the boundary, in the following, we consider the pre-implementability function $p = g[(N^+, N^-)]$ only constrained on the product of boundary $(\partial \mathcal{F})^2$.    

\section{Preorder on the resource space} \label{app: preorder}

    In the previous, we introduce the implementability function $p$ on the vector space $V$ and prove that it is a norm.
    With this norm, we can induce a topology on the space and other structures investigated in analysis, like Borel $\sigma$-algebra.
    However, the vector space $V$ we discussed is a subset of the whole space of resource $\mathcal{B}(X)$, and possibly a proper subset.
    For example, if the free set of dynamical resources consists of the local operation and classical channel (LOCC), the vector space $V(AB \rightarrow AB)$ is the proper space of $\mathcal{B}(AB \rightarrow AB)$, and typically, the CNOT gate is excluded from $V(AB \rightarrow AB)$~\cite{guo2023noise}.

    The resource $N \in V$ can be simulated by the free set with quasiprobability decomposition, while the resource $N \in \mathcal{B}\setminus V$ cannot.
    This means that the resource $N \in \mathcal{B}\setminus V$ is more precious than the resource $N \in V$ qualitatively.
    If $V = \mathcal{B}$, then all resources have the same quality, and the implementability norm $p$ can quantify the preciousness, for its physical meaning and mathematical properties.
    Otherwise, if $V \subsetneq \mathcal{B}$, we need other quantities to measure the resource $N \in \mathcal{B}\setminus V$ out of the space $V$.
    Thus, we consider the preorder relations of the resources, which should admit some physical meaning.
    The measures of resources should be an order-preserving map from the resources to a number.
    To construct the preorder relations, we start with the operations and its objects with physical meaning.

    For the operations on resources, we only assume the addition and scalar multiplication on the resources, because the multiplication of resources does not exist in the space of static resources on its own. 
    The scalar multiplication corresponds to the normalization of resources, which has less interesting physical results, and we persist it for a simple mathematical description.
    The addition and its inverse, which, combined with scalar multiplication, correspond to the convex combination and affine (vector) combination, have different physical meanings.
    The addition can be realized by the direct mixture of resources, while the subtraction requires a mixture with signatures.
    This difference leads to the implementability function discussed previously.
    For the objects of the operations, resource $N \in \mathcal{B}\setminus V$ is needed.
    Besides, we have two different sets, the extended free set $\mathcal{C} = \mathrm{conv}(\mathcal{F} \cup -\mathcal{F})$ and the space $V$.

    Then, we can define four preorder relations. 
    The first is that for $N, M \in \mathcal{B}$, $N \preceq M$ if $N \in \mathrm{conv}(\{M\} \cup \mathcal{C})$, where the term $\mathrm{conv}$ denotes the convex closure, and it induces a trivial equivalence relation $N \sim M \Leftrightarrow N = M$.
    The second is that $N \preceq M$ if $N \in \mathrm{conv}(\{M\} \cup V)$, and the induced equivalence relation is $N \sim M \Leftrightarrow N \in M + V$, which give a homomorphism $\phi: \mathcal{B} \rightarrow \mathcal{B}/V$ of resource space $\mathcal{B}$ with kernel $\ker \phi = V$.
    The third is that $N \preceq M$ if $N \in \langle\{M\} \cup \mathcal{C}\rangle$, which is the same as the fourth, $N \preceq M$ if $N \in \langle\{M\} \cup V\rangle$, and they induce an equivalence relation $N \sim M \Leftrightarrow N \in \langle \{M\} \cup \mathcal{F}\rangle \setminus V$.
    The equivalence induced from the third and fourth pre-order is so coarse that it cannot quantify the preciousness of elements in the equivalence class.
    Therefore, the second equivalence should be a suitable choice.

    Assume $q: \mathcal{B}/V \rightarrow \mathbb{R}$ is a measure of the equivalence class of resources on quotient space $\mathcal{B}/V$, then the function $P(N) = (q(N + V), p(N))$ is a measure of resources on $\mathcal{B}$, which admits the lexicographical order that $P(N) \preceq P(M)$ if $q(N + V) < q(M + V)$, or $q(N + V) = q(M + V)$ and $p(N) \leq p(M)$. 
    This lexicographical order reflects the fact that the resource $N \in \mathcal{B}\setminus V$ is more precious than the resource $N \in V$ qualitatively.
    One simplest measure $q$ can be selected as the $L_2$ norm on $\mathcal{B}/V$.
    For more interest choices of $q$, other structures, like the multiplication of resources, may be under consideration.

    {

\section{Probabilistic Error Cancellation with Noisy Pauli Basis} \label{app: noisy}

\subsection{Invertibility of noise map}
    In this appendix, we calculate the criterion of the invertibility of the noise map $\Theta$ in experimental finite precision, which is based on the probability inequality.
    
    Assume for each element $\Theta_{ij}$ of the linear map $\Theta$, $N$ times of single-shot measurements is employed.
    The variance of the element $\Theta_{ij}$ is $\mathrm{Var}\Theta_{ij} \leq \frac{1}{N}$, since $\sum_j \Theta_{ij} =1$, and $0<\Theta_{ij}<1$.
    For the determinant of the linear map $\Theta$, the difference is
    \begin{equation}
        \Delta \det \Theta = \sum_{i,j} A_{ij} \Delta \Theta_{ij},
    \end{equation}
    where $A_{ij}$ is the cofactor of $\Theta_{ij}$.
    Thus, the variance of the determinant of the linear map $\Theta$ is
    \begin{equation}
        \mathrm{Var} \det \Theta = \sum_{i, j} A_{ij}^2 \mathrm{Var}\Theta_{ij} \leq \frac{1}{N} \Vert \hat{A} \Vert_2^2,
    \end{equation}
    where $\hat{A}$ is the matrix of cofactors of $\Theta_{ij}$.
    With the Cramer's rules, the matrix of cofactors $\hat{A}$ satisfies
    \begin{equation}
        \hat{A}^{T} = \Theta^{-1} \det \Theta.
    \end{equation} 
    Therefore, the variance of the determinant of the linear map $\Theta$ is
    \begin{equation}
        \mathrm{Var} \det \Theta \leq \frac{\det^2 \Theta}{N} \Vert \Theta^{-1} \Vert_2^2,
    \end{equation}
    where $\Vert A \Vert_2 = \sqrt{\mathrm{Tr}A^{\dagger}A}$ is the Frobenius norm.

    For simplicity, denote the true value of the determinant of the linear map $\Theta$ as ${\det}_0$, and the measurement outcome as $\det$.
    By the Chebyshev's inequality, 
    \begin{equation}
        \mathbb{P}(\det\geq {\det}_0 + \epsilon) \leq \exp [-\Lambda^*(\epsilon)],
    \end{equation}
    where $\Lambda^*(\epsilon) = \sup_{\lambda}[\lambda \epsilon - \Lambda(\lambda)]$ is the Legendre transform of the cumulant generating function 
    \begin{equation}
        \Lambda(\lambda) = \log \mathbb{E}[\exp\lambda (\det-{\det}_0)]
    \end{equation}
    By Bennett's lemma~\cite{dembo2009large},  
    \begin{equation}
        \mathbb{E}(e^{\lambda x}) \leq e^{\lambda \bar{x}}\left[\frac{(b-\bar{x})^2}{(b-\bar{x})^2 + \sigma^2} e^{-\frac{\lambda \sigma^2}{b-\bar{x}}} + \frac{\sigma^2}{(b-\bar{x})^2 + \sigma^2} e^{\lambda (b-\bar{x})}\right],
    \end{equation}
    where $\mathrm{Var} x \leq \sigma^2$, and $x \leq b$.

    Since the absolute value of the determinant of the linear map $\Theta$ is the volume of the parallelepiped spanned by the columns or rows of $\Theta$, with the fact that $\sum_j \Theta_{ij} = 1$, and $\Theta_{ij}\geq0$, the maximum of absolute value of determinant $\det$ is the double of the volume of the regular $D$-simplex, where $D$ is the dimension of the matrix $\Theta$.
    Thus, $|\det| \leq 1$, and $|\det-\det^*| \leq 2$.
    Then, the generating function of the determinant of the linear map $\Theta$ is
    \begin{align}
        \mathbb{E}e^{\lambda (\det-{\det}_0)} & \leq e^{2 \lambda} \left[\frac{4}{4 + \sigma^2} e^{-(\frac{\sigma^2}{2}+2)\lambda} + \frac{\sigma^2}{4 + \sigma^2} \right] \nonumber\\
        & \leq e^{2 \lambda} \left[\frac{4}{4 + \sigma^2} e^{-2\sigma\lambda} + \frac{\sigma^2}{4 + \sigma^2} \right]
    \end{align}
    where $\sigma^2 = \frac{\det^2 \Theta}{N} \Vert \Theta^{-1} \Vert_2^2$.
    The cumulant generating function is
    \begin{equation}
        \Lambda(\lambda) \leq 2 \lambda +  \log \left[1-\gamma + \gamma e^{-2\sigma\lambda}\right] \equiv f(\lambda),
    \end{equation}
    where $\gamma = \frac{\sigma^2}{4 + \sigma^2}$.
    Direct calculation shows that $f(0) = 0, f'(0)=0$ and 
    \begin{equation}
        f''(x) = \frac{(2\sigma)^2(1-\gamma)\gamma e^{-2\sigma\lambda}}{\left(1-\gamma + \gamma e^{-2\sigma\lambda}\right)^2} \leq \sigma^2.
    \end{equation}
    By the Taylor's theorem, we have
    \begin{equation}
        f(\lambda) = f(0) + f'(0) \lambda + \frac{1}{2} f''(\xi) \lambda^2 \leq \frac{\sigma^2}{2} \lambda^2,
    \end{equation}
    where $\xi \in [0, \lambda]$.
    and the Legendre transform is
    \begin{equation}
        \Lambda^*(\epsilon) \geq \sup_{\lambda}\left[\lambda \epsilon - \frac{\sigma^2}{2} \lambda^2\right] = \frac{\epsilon^2}{2\sigma^2}  .
    \end{equation}

    Therefore, the probability that the true value of the determinant of the linear map $\Theta$ is not invertible is
    \begin{align}
        \mathbb{P}({\det}_0 = 0) &\leq \mathbb{P}(\det \geq {\det}_0 + \det) \nonumber\\
        & \leq \exp\left[-\frac{\det^2}{2\sigma^2}\right] \nonumber\\
        & = \exp\left[-\frac{N}{2\Vert \Theta^{-1} \Vert_2^2}\right].
    \end{align}
    With the probability $\delta$ of error tolerance, we have
    \begin{equation}
        \mathbb{P}({\det}_0 = 0) \leq \exp\left[-\frac{N}{2\Vert \Theta^{-1} \Vert_2^2}\right] \leq \delta,
    \end{equation}
    which is equivalent to
    \begin{equation}
        \Vert \Theta^{-1} \Vert_2 \leq \sqrt{\frac{N}{2 \log \frac{1}{\delta}}}.
    \end{equation}
    Moreover, since the matrix algebra is Banach algebra with respect to the $\Vert \cdot\Vert_2$, 
    \begin{equation}
        \Vert \Theta \Vert_2 \cdot \Vert \Theta^{-1} \Vert_2 \geq \Vert \hat{I} \Vert_2 = \sqrt{D},
    \end{equation}
    Therefore, for the given number $N$ of measurements, if the matrix $\Theta$ satisfies that 
    \begin{equation}
        \Vert \Theta \Vert_2 \geq \sqrt{\frac{2 D \log \frac{1}{\delta}}{N}},
    \end{equation}
    where $D$ is the dimension of the matrix $\Theta$, the linear map $\Theta$ is invertible with the probability $1 - \delta$.

    \subsection{Bias of imperfect cancellation}

    Without mitigating the noise in the simulation of inverse noise operation $\mathcal{E}^{-1}$, we would like to estimate the distance between the noisy mitigated error channel $\mathcal{E}_{\lambda}^{-1} \circ \mathcal{E} = \Theta(\mathcal{E}^{-1}) \circ \mathcal{E}$ and the identical channel $\mathcal{I}$.
    The bias of the expectation is 
    \begin{align}
        \delta_{\lambda} & = |\mathrm{Tr}[O\mathcal{U}(\rho)] - \mathrm{Tr}[O\mathcal{E}_{\lambda}^{-1} \circ \mathcal{E}\circ \mathcal{U} (\rho)]| \nonumber\\
        & \leq \Vert\mathcal{U}(\rho) - \mathcal{E}_{\lambda}^{-1} \circ \mathcal{E}\circ \mathcal{U}(\rho)\Vert_1 \nonumber\\
        & = p_{\mathcal{Q}(A)}\left[\mathcal{U}(\rho) - \mathcal{E}_{\lambda}^{-1} \circ \mathcal{E}\circ \mathcal{U}(\rho) \right] \nonumber \\
        & \leq \Vert\mathcal{U} - \mathcal{E}_{\lambda}^{-1} \circ \mathcal{E}\circ \mathcal{U} \Vert_{\mathcal{\diamond}\mathcal{Q}(A)} \nonumber\\
        & \leq p_{\mathcal{Q}(A\rightarrow A)}\left(\mathcal{U} - \mathcal{E}_{\lambda}^{-1} \circ \mathcal{E}\circ \mathcal{U} \right) \nonumber\\
        & = p_{\mathcal{Q}(A\rightarrow A)}\left(\mathcal{I} - \mathcal{E}_{\lambda}^{-1} \circ \mathcal{E} \right),
    \end{align}
    where we use $\Vert \cdot \Vert_1 = p_{\mathcal{Q}(A)}$ in Proposition~\ref{pro: trace_norm}, $\Vert \cdot \Vert_{\diamond\mathcal{Q}(A)}$ is the norm induced by $p_{\mathcal{Q}(A)}$ in Definition~\ref{def: induced_norm}, and the last equality holds for the unitary operation $\mathcal{U}$.
    For simplicity, we denote $\mathcal{Q} = \mathcal{Q}(A\rightarrow A)$ in the following.
    With $\mathcal{E}_{\lambda}^{-1} = \Theta(\mathcal{E}^{-1})$ we have
    \begin{align}
        & p_{\mathcal{Q}}\left(\mathcal{I} - \mathcal{E}_{\lambda}^{-1} \circ \mathcal{E} \right)
        \leq p_{\mathcal{Q}}\left(\Theta(\mathcal{E}^{-1}) - \mathcal{E}^{-1}\right) p_{\mathcal{Q}}(\mathcal{E})  \\
        & = p_{\mathcal{Q}}\left(\Theta(\mathcal{E}^{-1}) - \mathcal{E}^{-1}\right) \leq \Vert\Theta - \mathrm{id}\Vert_{\mathcal{Q}} p_{\Theta(\mathcal{Q})}(\mathcal{E}_{\lambda}^{-1}), \nonumber
    \end{align}
    where $\Vert\Omega \Vert_{\mathcal{Q}} = \max \frac{p_{\mathcal{Q}}(\Omega(\mathcal{N}))}{p_{\mathcal{Q}}(\mathcal{N})}$, with the maximum being over all Pauli diagonal operations.

    For the Pauli diagonal error, $\mathcal{N} = \sum_i n_i \mathcal{P}_i$, we have 
    \begin{align}
        \frac{p_{\mathcal{Q}}(\Omega(\mathcal{N}))}{p_{\mathcal{Q}}(\mathcal{N})} & = \frac{p_{\mathcal{Q}}\left(\sum_{i,j} n_i \Omega_{ij} \mathcal{P}_j\right)}{p_{\mathcal{Q}}\left(\sum_{i} n_i \mathcal{P}_j\right)} \nonumber \\
        & = \frac{\sum_{i,j} |n_i \Omega_{ij}|}{\sum_{i} |n_i|} \leq \max_i \sum_{j} |\Omega_{ij}|,
    \end{align}
    and for $\mathcal{N} = \mathcal{P}_{i_0}$, where $i_0$ reaches the maximum, the equality holds:
    \begin{equation}
        \Vert\Omega \Vert_{\mathcal{Q}} = \max_i \sum_{j} |\Omega_{ij}|.
    \end{equation}
    Obviously, the noisy realizations of Pauli basis $\mathcal{K}_i = \Theta(\mathcal{P}_i)$ are also physical, so $\Theta_{ij} \geq 0$, and $\sum_{j} \Theta_{ij} = 1$.
    We thus have
    \begin{equation}
        \Vert\Theta - \mathrm{id}\Vert_{\mathcal{Q}} = 1 + \max_i \left(\sum_{j \neq i} \Theta_{ij} - \Theta_{ii}\right)\leq 2 \Theta_{\lambda},
    \end{equation}
    where $\Theta_{\lambda} = 1 - \min_i \Theta_{ii}$ is the maximal error probability of noisy Pauli gates.
    Thus, the bias is bounded by
    \begin{equation}
        \delta_{\lambda} \leq 2 \Theta_{\lambda} p_{\Theta(\mathcal{Q})}(\mathcal{E}_{\lambda}^{-1}).
    \end{equation}        
    It is saturated, when $\mathcal{E} = \mathcal{P}_{i_0}$, where $\Theta_{ii}$ reaches its minimum at $i_0$.
    Thus, the bound is (asymptotically) tight over all the possible error channels.
    This result allows for estimating the upper bound of the imperfection in experiments, with the cost of simulation $p_{\Theta(\mathcal{Q})}(\mathcal{E}_{\lambda}^{-1})$ and the calibration of Pauli gates $\Theta_{\lambda}$.

    For the cancellation of error of many layers $\mathcal{U} = \prod \circ \mathcal{L}_i$, there are the direct and separate way to cancel the error.
    Let the noisy realization of circuit be $\mathcal{U}_{\lambda} = \overleftarrow{\bigcirc}_{i = 1}^{L} \mathcal{L}_{i\lambda}$, where $\mathcal{L}_{i\lambda} = \mathcal{E}_i \circ \mathcal{L}_i$, then the error channel of the circuit is 
    \begin{equation}
        \mathcal{E} = \overleftarrow{\bigcirc}_{i = 1}^{L} \mathcal{\tilde{E}}_i,
    \end{equation}
    where $\mathcal{\tilde{E}}_i = \left( \overleftarrow{\bigcirc}_{j > i}^{L} \mathcal{L}_{j} \right) \circ \mathcal{E}_i \circ \left(\overrightarrow{\bigcirc}_{j > i}^{L} \mathcal{L}_{j}^{\dagger}\right)$.
    Here, the arrow above the symbol $\bigcirc$ represents the acting direction of layers. 

    For the separate cancellation method, the noisy realization of the circuit with the PEC method is 
    \begin{equation}
        \mathcal{U}_{\mathrm{PEC}} = \overleftarrow{\bigcirc}_{i = 1}^{L} \mathcal{L}_{i\mathrm{PEC}},
    \end{equation}
    where $\mathcal{L}_{i\mathrm{PEC}} = \mathcal{E}_{i\lambda}^{-1} \circ \mathcal{E}_i \circ \mathcal{L}_i$.
    The noisy mitigated error channel is
    \begin{equation}
        \mathcal{E}_{\mathrm{S}}^{-1} \circ \mathcal{E} = \overleftarrow{\bigcirc}_{i = 1}^{L} \mathcal{\tilde{E}}_{i\mathrm{PEC}},
    \end{equation}
    where $\mathcal{\tilde{E}}_{i\mathrm{PEC}} = \left( \overleftarrow{\bigcirc}_{j > i}^{L} \mathcal{L}_{j} \right) \circ \mathcal{E}_{i\lambda}^{-1} \circ \mathcal{E}_i \circ \left(\overrightarrow{\bigcirc}_{j > i}^{L} \mathcal{L}_{j}^{\dagger}\right)$.
    The bias of the noisy mitigated error channel $\mathcal{E}_{\mathrm{S}}^{-1} \circ \mathcal{E} = \Theta(\mathcal{E}^{-1}) \circ \mathcal{E}$ is 
    \begin{align}
        & \delta_{\lambda\mathrm{S}} \leq p_{\mathcal{Q}}\left(\overleftarrow{\bigcirc}_{i = 1}^{L} \mathcal{\tilde{E}}_{i\mathrm{PEC}} - \mathcal{I}\right)  \nonumber \\
        &~~~~~~~~\leq \sum_{j=1}^{L} p_{\mathcal{Q}}\left(\overleftarrow{\bigcirc}_{i = 1}^{j} \mathcal{\tilde{E}}_{i\mathrm{PEC}} - \overleftarrow{\bigcirc}_{i = 1}^{j-1} \mathcal{\tilde{E}}_{i\mathrm{PEC}}\right),
    \end{align}
    where the triangle inequality for the implementability function $p_{\mathcal{Q}}$ is used.
    By using the sub-multiplicity, Eq.~(\ref{eq: sub-multiplicity}), we have 
    \begin{align}
        & p_{\mathcal{Q}}\left(\overleftarrow{\bigcirc}_{i = 1}^{j} \mathcal{\tilde{E}}_{i\mathrm{PEC}} - \overleftarrow{\bigcirc}_{i = 1}^{j-1} \mathcal{\tilde{E}}_{i\mathrm{PEC}}\right) \nonumber \\
        &~~~~ \leq p_{\mathcal{Q}}\left(\mathcal{\tilde{E}}_{j\mathrm{PEC}} - \mathcal{I}\right) p_{\mathcal{Q}}\left(\overleftarrow{\bigcirc}_{i = 1}^{j-1} \mathcal{\tilde{E}}_{i\mathrm{PEC}}\right) \nonumber \\
        &~~~~ \leq p_{\mathcal{Q}}\left(\mathcal{E}_{j\lambda}^{-1} - \mathcal{E}_{j}^{-1}\right) \prod_{i = 1}^{j-1} p_{\mathcal{Q}}\left(\mathcal{E}_{i\lambda}^{-1}\right) \nonumber \\
        &~~~~ \leq 2 \Theta_{\lambda} \prod_{i = 1}^{j} p_{\Theta(\mathcal{Q})}\left(\mathcal{E}_{i\lambda}^{-1}\right),
    \end{align}
    so the bias is bounded by 
    \begin{equation}
        \delta_{\lambda\mathrm{S}} \leq 2 \Theta_{\lambda} \sum_{j=1}^{L} \prod_{i = 1}^{j} p_{\Theta(\mathcal{Q})}\left(\mathcal{E}_{i\lambda}^{-1}\right).
    \end{equation}

    For the noise map $\Theta$ preserves the set $\mathcal{E}^{-1} \circ \mathcal{Q}$, the error of noisy cancellation for each layer in the separate cancellation method is also a CPTP quantum channel, and the bias is bounded by
    \begin{equation}
        p_{\mathcal{Q}}\left(\mathcal{I} - \mathcal{E}_{\lambda}^{-1} \circ \mathcal{E} \right) \leq 2(1 - \nu_0),
    \end{equation}
    where $\nu_0$ is the component of $\mathcal{I}$ in $\mathcal{E}_{\lambda}^{-1} \circ \mathcal{E}$.
    For the Pauli diagonal operation $\mathcal{E}$, it has the Lindblad representation
    \begin{equation}
        \mathcal{E} = e^{\mathcal{L}(\lambda_i)},
    \end{equation}
    where $\mathcal{L}(\lambda_i) = \sum_{i}\lambda_i (\mathcal{P} - \mathcal{I})$.
    Constrained to the subalgebra of Pauli diagonal operations, it is an Abelian algebra with the simple multiplication rule 
    \begin{equation}
        e^{\mathcal{L}(\lambda_i)}e^{\mathcal{L}(\lambda_i')} = e^{\mathcal{L}(\lambda_i + \lambda_i')}.
    \end{equation}
    The component of $\mathcal{I}$ for $e^{\mathcal{L}(\lambda_i)}$
    For a CPTP Pauli-Lindblad channel, $\lambda_i>0$,
    \begin{equation}
        e^{\mathcal{L}(\lambda_i)} = \bigcirc_{i} (\omega_i\mathcal{I} + (1-\omega_i) \mathcal{P}_i),
    \end{equation}
    where $\omega_i = \frac{1}{2}(1 + e^{-2\lambda_i})>0$, the component of $\mathcal{I}$
    \begin{equation}
        \nu_0 \geq \prod_i \omega_i \geq e^{-\sum_i \lambda_i} = \frac{1}{\sqrt{p_{\mathcal{Q}}(e^{-\mathcal{L}(\lambda_i)})}}, 
    \end{equation}
    by convexity of the function $e^{x}$.
    Therefore, the bias of is bounded by
    \begin{equation}
        \delta_{\lambda\mathrm{S}} \leq 2\left(1 - \frac{1}{\prod_{i=1}^L p_{\mathcal{Q}}^{1/2}[(\mathcal{E}_{i\lambda}^{-1})^{-1} \circ \mathcal{E}_i^{-1}]}\right) \leq 2.
    \end{equation}

    Now, We denote $\Theta = \mathrm{id} + \tilde{\Theta}$.
    The inverse of $\mathcal{E}_{i\lambda}^{-1}$ is
    \begin{align}
        (\mathcal{E}_{i\lambda}^{-1})^{-1} & = (\mathcal{E}_i^{-1} + \tilde{\Theta}(\mathcal{E}_i^{-1}))^{-1} \\
        & = \mathcal{E}_i \circ \left(\sum_n [-\mathcal{E}_i\circ \tilde{\Theta}(\mathcal{E}_i^{-1})]^n\right). \nonumber
    \end{align}
    The implementability function is
    \begin{align}
        p_{\mathcal{Q}}[(\mathcal{E}_{i\lambda}^{-1})^{-1} \circ \mathcal{E}_i^{-1}] & \leq \sum_n p_{\mathcal{Q}}^n[\mathcal{E}_i\circ \tilde{\Theta}(\mathcal{E}_i^{-1})] \nonumber \\
        & = \frac{1}{1 - p_{\mathcal{Q}}[\mathcal{E}_i\circ \tilde{\Theta}(\mathcal{E}_i^{-1})]},
    \end{align}
    where
    \begin{equation}
        p_{\mathcal{Q}}[\mathcal{E}_i\circ \tilde{\Theta}(\mathcal{E}_i^{-1})]  
        \leq p_{\mathcal{Q}}[\mathcal{I} - \mathcal{E}_i\circ \mathcal{E}_{i\lambda}^{-1}] \leq 2 \Theta_{\lambda},
    \end{equation}
    with the assumption that $\mathcal{E}_i\circ \mathcal{E}_{i\lambda}^{-1}$ is a CPTP channel.
    Thus, the bias is bounded by
    \begin{equation}
        \delta_{\lambda\mathrm{S}} \leq 2(1 - (1-2\Theta_{\lambda})^{L/2}).
    \end{equation}

    {
    \subsection{Bias of imperfect error model}

        The bias of the expectation of Pauli operator $\hat{O}$ is bounded by 
        \begin{align}
            \delta_O & \leq p_{\mathcal{Q}}(\mathcal{I} - \mathcal{E}^{-1} \circ \hat{\mathcal{E}}) \nonumber \\
            &= p_{\mathcal{Q}}(\mathcal{E}^{-1} \circ \hat{\mathcal{E}}) +|1-\nu_0| - |\nu_0|,
        \end{align}
        where $\mathcal{E}^{-1} \circ \hat{\mathcal{E}} = e^{\mathcal{L}(\Delta \lambda_i)}$.
        For a Pauli diagonal operation $e^{\mathcal{L}(\lambda_i)}$, its component in $\mathcal{I}$ is
        \begin{align}
            \nu_0 & =\frac{1}{4^n} \sum_k e^{-2\sum_i M_{ki} \lambda_i}\geq e^{-\frac{2}{4^n}\sum_{i,k}M_{ki} \lambda_i} \nonumber \\
            & = e^{-\sum_{i} \lambda_i} \geq 0,
        \end{align}
        where $M_{ki} = 1$ if $\hat{P}_i$  anti-commutes with $\hat{P}_k$, and $M_{ki} = 0$ otherwise.
        The inequality holds for the convexity of the exponential function $e^{x}$, and the second equality holds for any $\hat{P}_i \neq \hat{I}$. There are half the number of $\hat{P}_j$ anti-commuting with $\hat{P}_k$, and $\lambda_0 \equiv 0$ since 
        \begin{equation}
            \mathcal{L}_{0}(\lambda_0) = \lambda_0 (\mathcal{I} -\mathcal{I}) \equiv 0.
        \end{equation}
        Therefore, the distance in the implementability function is 
        \begin{equation}
            p_{\mathcal{Q}}\left(\mathcal{I} - \mathcal{E}_{\lambda}^{-1} \circ \mathcal{E} \right) = p_{\mathcal{Q}}\left(\mathcal{E}_{\lambda}^{-1} \circ \mathcal{E} \right) + |1 - \nu_0| - \nu_0.
        \end{equation}
        Furthermore, the implementability function of the Pauli diagonal operation $e^{\mathcal{L}(\lambda_i)}$ is 
        \begin{align}
            p_{\mathcal{Q}}\left(\exp {\mathcal{L}(\lambda_i)}\right) & \leq \prod_i p_{\mathcal{Q}}(e^{\lambda_i (\mathcal{P}_i - \mathcal{I})}) \nonumber \\
            & = e^{-2 \sum_i \min\{\lambda_i,0\}},
        \end{align}  
        and
        \begin{align}
            &|1-\nu_0| - \nu_0  = \left\{\begin{array}{lc}
            -1, & \nu_0 >1, \\
            1-2 \nu_0, & 0<\nu_0 \leq 1, \\
            \end{array}\right. \\
            & = 1-2 \min\{\nu_0,1\} 
            \leq 1 - 2 e^{-\max\{\sum_i \lambda_i ,0\}}. \nonumber 
        \end{align}
    
        If the error channel $\mathcal{E}^{-1} \circ \hat{\mathcal{E}}$ is CPTP, $\Delta \lambda_i > 0, p_{\mathcal{Q}}(\mathcal{E}^{-1} \circ \hat{\mathcal{E}}) = 1$, and the upper bound is 
        \begin{equation}
            \delta_{O} \leq 2(1-\nu_0) \leq 2\left(1 - e^{-\sum_i \Delta\lambda_i}\right).
        \end{equation}
        If the error channel $\mathcal{E}^{-1} \circ \hat{\mathcal{E}}$ is not CPTP,  the upper bound is 
        \begin{align}
            \delta_{O} & \leq p_{\mathcal{Q}}(\mathcal{E}^{-1} \circ \hat{\mathcal{E}}) +|1-\nu_0| - \nu_0 \\
            & \leq e^{-2 \sum_i \min\{\Delta\lambda_i,0\}} - 2 e^{-\max\{\sum_i \Delta\lambda_i ,0\}} + 1. \nonumber
        \end{align}  

    }

    Then, we consider the non-Pauli-diagonal error $\mathcal{E}$.
    Since the Pauli operators form a complete set of the bases for the $U(2^n)$ group, it can be expanded as 
    \begin{equation}
        \mathcal{E}(\cdots) = \sum_{i,j} \omega_{ij} P_i (\cdots) P_j.
    \end{equation}
    By using the Pauli twirling technique, the twirled error can be expressed as 
    \begin{align}
        \hat{\mathcal{E}} & = \mathbb{E}_k \mathcal{P}_k \circ \mathcal{E} \circ \mathcal{P}_k(\cdots)
        = \sum_{i,j} \omega_{ij} \mathbb{E}_k P_k P_i P_k (\cdots) P_k P_j P_k \nonumber \\
        & = \sum_{i,j} \omega_{ij} \mathbb{E}_k (\epsilon_{ik} \epsilon_{jk}) P_i (\cdots) P_j 
        = \sum_{i,j} \omega_{ij} \delta_{ij}  P_i (\cdots) P_j \nonumber \\
        & = \sum_{i} \omega_{ii} P_i (\cdots) P_i,
    \end{align}
    where $\epsilon_{ik} = 1$ if $P_i$ commutes with $P_k$, otherwise $\epsilon_{ik} = -1$, and $\mathbb{E}_k (\epsilon_{ik} \epsilon_{jk}) = \delta_{ij}$ can be verified easily.

    For the worst case, no random compiling is employed in the Pauli twirling technique, the canceled error is written as
    \begin{equation}
        \hat{\mathcal{E}}^{-1} \circ \mathcal{E} = \mathcal{I} + \hat{\mathcal{E}}_0^{-1} \circ \Delta \mathcal{E},
    \end{equation}
    where $\Delta \mathcal{E} = \sum_{i\neq j} \omega_{ij} \delta_{ij}  P_i (\cdot) P_j$ is not trace-preserving.
    Therefore, the upper bound is obtained as
    \begin{equation}
        \delta_O \leq p_{\mathcal{Q}}(\hat{\mathcal{E}}^{-1} \circ \Delta \mathcal{E}).
    \end{equation}
    For the separate cancellation method, the canceled error for $L$ layers is expressed as 
    \begin{equation}
        (\hat{\mathcal{E}}^{-1} \circ \mathcal{E})^L = \mathcal{I} + \sum_{l = 1}^L C_L^l (\hat{\mathcal{E}}_0^{-1} \circ \Delta \mathcal{E})^l,
    \end{equation} 
    and the upper bound for the bias is calculated as
    \begin{equation}
        \delta_{O} \leq \sum_{l = 1}^L C_L^l p_{\mathcal{Q}}(\hat{\mathcal{E}}^{-1} \circ \Delta \mathcal{E})^l = [1 + p_{\mathcal{Q}}(\hat{\mathcal{E}}^{-1} \circ \Delta \mathcal{E})]^L - 1,
    \end{equation}

    Although we could not explicitly calculate $p_{\mathcal{Q}}(\hat{\mathcal{E}}^{-1} \circ \Delta \mathcal{E})$ for $\omega_{ij} \neq \omega_{ii} \delta_{ij}$,  we obviously have $p_{\mathcal{Q}}(\hat{\mathcal{E}}^{-1} \circ \Delta \mathcal{E}) \rightarrow 0$ when $\omega_{ij} \rightarrow \delta_{ij}$.
    We assume that the Pauli twirling technique is employed with $N$ shots. According to the strong large number theorem, the imperfect twirled error model is expressed as
    \begin{equation}
        \tilde{\mathcal{E}}(\cdots) = \sum_{i, j} {\omega}_{ij} x_{ij} P_i (\cdots) P_j,
    \end{equation}
    where $x_{ij} \sim N(\delta_{ij}, \sigma_N)$ follows a Gaussian distribution, and $\sigma_N = \sigma/\sqrt{N}$ for some finite derivation $\sigma$.
    Since $p_{\mathcal{Q}}(\cdots) = \Vert \cdots \Vert_{\diamond}$ is a norm on the Hermitian-preserving bounded operators, isomorphic to their Choi matrices, and the topology induced by the norms of matrix algebra are all equivalent, we can expect the concentration inequality of probability measure as
    \begin{equation}
        \mathbb{P}(\Vert \hat{\mathcal{E}}^{-1} \circ \Delta \tilde{\mathcal{E}}\Vert_{\diamond} \geq \epsilon) \leq Ce^{-N c\epsilon^2}
    \end{equation}
    for some $c, C > 0$.
    Therefore, for $\delta \ll 1$, we have 
    \begin{equation}
        \mathbb{P}(\delta_{O} \geq \delta) \leq Ce^{-N c[(1 + \delta)^{1/L} - 1]^2} \approx Ce^{-N c\delta^2/L^2}.
    \end{equation}
    Then, given a tolerant precision $\delta \ll 1$ for the estimation and a tolerant probability $\gamma \ll 1$ for the failure, the number of shots  $N$  should satisfies
    \begin{equation}
        N \gtrsim \frac{L^2}{\delta^2} (A - B \log\gamma),
    \end{equation}
    where $A, B \geq 0$ are some constants depending on the parameters $\omega_{ij}$ of the error model.

\section{Details of Numerical Simulation} \label{app: simulation}

    \subsection{Simulation of noisy cancellation}

    A $2$-qubit system is simulated.
    For simplicity, the identical channel $\mathcal{I}$ is selected as the ideal circuit.
    The single-layer error channel $\mathcal{E}_0$ is a Pauli-Lindblad error, where the parameters $\lambda_i$ are random sampled with fixed error rate $\lambda=\sum_i \lambda_i$.
    The inverse of the error $\mathcal{E}_0^{-1}$ is the Pauli-Lindblad channel with the parameters $-\lambda_i$.
    The noises $\mathcal{N}_i$ of the Pauli gates are also randomly sampled Pauli-Lindblad errors, with the same error rate $\lambda$ as the error channel.
    The noise map $\Theta$ is depicted by the noises $\mathcal{N}_i$, the action of  $\Theta$ on the error is 
    \begin{equation}
        \mathcal{E} = \sum_i x_i \mathcal{P}_i  \mapsto \Theta(\mathcal{E}) = \sum_i x_i \mathcal{N}_i \circ \mathcal{P}_i.
    \end{equation}

    The error of separate cancellation method for $L$-layer circuit is the $L$-times composition of the noisy cancelled error $\mathcal{E}_0^{-1} \circ \Theta(\mathcal{E}_0)$, i.e.
    \begin{equation}
        \mathcal{E}_{\mathrm{S}} = [\Theta(\mathcal{E}_0^{-1}) \circ \mathcal{E}_0]^L.
    \end{equation}
    The error of direct cancellation method for $L$-layer circuit is the noisy cancelled error of $L$-layer error $\mathcal{E}_0^L$, i.e.
    \begin{equation}
        \mathcal{E}_{\mathrm{D}} = \Theta(\mathcal{E}_0^{-L}) \circ \mathcal{E}_0^L.
    \end{equation}

    For the expectations of Pauli operators, the ideal expectation is 
    \begin{equation}
        \braket{\hat{P}_i}_0 = \mathrm{Tr}[\hat{P}_i\rho],
    \end{equation}
    and the noisy expectation is 
    \begin{align}
        \braket{\hat{P}_i} &= \mathrm{Tr}[\hat{P}_i\mathcal{E}_{\mathrm{S},\mathrm{D}}(\rho)] 
        = \mathrm{Tr}[\mathcal{E}_{\mathrm{S},\mathrm{D}}(\hat{P}_i)\rho] \nonumber \\
        & = \chi_{i\mathrm{S},\mathrm{D}}\braket{\hat{P}_i}_0,
    \end{align}
    where $\chi_{i\mathrm{S},\mathrm{D}}$ is the eigenvalue of the Pauli diagonal error $\mathcal{E}_{\mathrm{S},\mathrm{D}}$ with respect to the Pauli operator $\hat{P}_i$, since the Pauli operator is the eigen-operator of Pauli diagonal operations.
    The bias of Pauli operator $\hat{P}_i$ expectation for noisy cancellation is $\delta_{\lambda \hat{P}_i} = |1-\chi_{i\mathrm{S},\mathrm{D}}|$.
    The biases in implementability function are $p_{\mathcal{Q}}(\mathcal{I} - \mathcal{E}_{\mathrm{S},\mathrm{D}})$, which is the $1$-norm with respect to the decomposition in the Pauli basis.
    The upper bounds of the bias with CPTP noisy cancelled error are $2[1-(1-2\Theta_{\lambda})^{L/2}]$ for separate cancellation method and $2\Theta_{\lambda}$ for direct cancellation method, where
    \begin{equation}
        \Theta_{\lambda} = 1 - \min \Theta_{ii} = 1 - \min \nu_{0i}.
    \end{equation}
    Here, $\nu_{0i}$ is the component of the identical operation $\mathcal{I}$ of the noise $\mathcal{N}_i$.

    \subsection{Simulation of imperfect error model}

    The method os simulation is similar to the simulation of noisy cancellation.
    The single-layer under-mitigated error $\hat{\mathcal{E}}_0^{-1} \circ \mathcal{E}_{0} = e^{\mathcal{L}(\Delta\lambda_i^{\mathrm{under}})}$ is a Pauli-Lindblad error with the random selected parameters $\Delta\lambda_i^{\mathrm{under}}>0$ and $\sum_i \Delta\lambda_i^{\mathrm{under}} = 0.05$.
    The single-layer over-mitigated error $\hat{\mathcal{E}}_0^{-1} \circ \mathcal{E}_{0} = e^{\mathcal{L}(\Delta\lambda_i^{\mathrm{over}})}$ is a Pauli-Lindblad error with the parameters $\Delta\lambda_i^{\mathrm{over}} = -\Delta\lambda_i^{\mathrm{under}}$.
    The total mitigated error of the circuit is
    \begin{equation}
        \hat{\mathcal{E}}^{-1} \circ \mathcal{E} = [\hat{\mathcal{E}}_0^{-1} \circ \mathcal{E}_{0}]^L.
    \end{equation}

\section{Relation to Other Quantities} \label{app: relation}

    \subsection{Diamond norm}
    
    On the operation algebra $\mathcal{B}(A \rightarrow A)$, we can induce a norm the implementability function $p_{\mathcal{F}(A)}$ on space of state $\mathcal{B}(A)$.
    \begin{definition} \label{def: induced_norm}
        The generalized diamond norm, with respect to the set of free states $\mathcal{F}(A)$, is defined as
        \begin{equation}
            \Vert \mathcal{N}\Vert_{\diamond \mathcal{F}(A)} \equiv \max \frac{p_{\mathcal{F}(A)}(\mathcal{N}(\rho))}{p_{\mathcal{F}(A)}(\rho)}.
        \end{equation}
    \end{definition}
    From the sub-multiplicativity of the implementability function, we have the following proposition.
    \begin{proposition} \label{pro: diamond}
        The generalized diamond norm is upper bounded by the implementability function with respect to the set of free operations $\mathcal{F}(A\rightarrow A)$,
        \begin{equation} 
            \Vert \mathcal{N}\Vert_{\diamond\mathcal{F}(A)} \leq p_{\mathcal{F}(A \rightarrow A)}(\mathcal{N}).
        \end{equation}
        The equality holds if and only if the free set of operation $\mathcal{F}(A\rightarrow A)$ is the resource nongenerating (RNG) set of operations $\mathcal{F}_{\max}(A\rightarrow A)$, i.e. the maximal assignment of operation for the free set $\mathcal{F}(A)$ of state
        \begin{equation}
            \Vert \mathcal{N}\Vert_{\diamond\mathcal{F}(A)} = p_{\mathcal{F}_{\max}(A \rightarrow A)}(\mathcal{N}).
        \end{equation}
    \end{proposition}
    \begin{proof}
        Proposition~\ref{pro: sub-multiplicity} shows the inequality.
        For the norm $\Vert\cdot \Vert_{\diamond \mathcal{F}(A)}$, there is a convex set $\mathcal{C}_{\diamond}(A\rightarrow A) = \{\Vert\mathcal{N} \Vert_{\diamond \mathcal{F}(A)}\leq 1\}$, where it is the Minkowski functional $\Vert\cdot \Vert_{\diamond \mathcal{F}(A)} = p_{\mathcal{C}_{\diamond}(A\rightarrow A)}$ of the convex set
        \begin{equation}
            \mathcal{C}_{\diamond}(A\rightarrow A) = \{\mathcal{N}: \Vert\mathcal{N} \Vert_{\diamond \mathcal{F}(A)}\leq 1\}.
        \end{equation}
        In particular, with the normalization of operations, $\mathcal{C}_{\diamond} = \mathrm{conv}(\mathcal{F}_{\diamond}\cup -\mathcal{F}_{\diamond}) $, where
        \begin{equation}
            \mathcal{F}_{\diamond}(A\rightarrow A) = \{\mathcal{N}: \Vert\mathcal{N} \Vert_{\diamond \mathcal{F}(A)} = 1\},
        \end{equation}
        The set $\mathcal{F}_{\diamond}(A\rightarrow A)$ is just the RNG set of operations, which is directly followed from the definition of the norm $\Vert\cdot \Vert_{\diamond \mathcal{F}}$.
    \end{proof}
    \noindent
    This induced norm can be interpreted as a generalization of the diamond norm
    \begin{equation}
        \Vert \mathcal{N}\Vert_{\diamond} \equiv \max_{\rho \in \mathcal{Q}(A \otimes A)} \frac{\Vert \mathrm{id} \otimes \mathcal{N}(\rho)\Vert_1}{\Vert \rho\Vert_1},
    \end{equation}
    since the trace norm $\Vert \cdot \Vert_1 = p_{\mathcal{Q}(A\otimes A)}$ (see Proposition~\ref{pro: trace_norm}).
    \begin{corollary}
        The diamond norm is the exponential of physical implementability 
        \begin{equation}
            \Vert \mathcal{N}\Vert_{\diamond} = p_{\mathcal{Q}(A \rightarrow A)}(\mathcal{N}).
        \end{equation}
    \end{corollary}
    \begin{proof}
        With Proposition~\ref{pro: diamond}, the equality holds $\Vert \mathcal{N}\Vert_{\diamond} = p_{\mathcal{F}_{\diamond}(A \rightarrow A)}(\mathcal{N})$
        only when 
        \begin{align}
            \mathcal{F}_{\diamond}(A \rightarrow A) & = \{\mathcal{N}: \forall \Vert \rho\Vert_1 = 1, \Vert \mathrm{id} \otimes \mathcal{N}(\rho)\Vert_1 = 1 \} \nonumber\\
            & = \{\mathcal{N}: \mathrm{id} \otimes \mathcal{N}\in \mathbb{B}(A\otimes A) \cap \mathrm{PTP}\}
        \end{align}
        is the set of operations $\mathcal{N}$ whose extension $\mathrm{id} \otimes \mathcal{N}$ is positive and trace preserving (PTP).
        This means that $\mathcal{N} \in \mathcal{Q}(A\rightarrow A)$ is CPTP, and $\mathcal{F}_{\diamond}(A \rightarrow A) = \mathcal{Q}(A\rightarrow A)$. 
    \end{proof}\noindent
    This result has also been proved in Theorem~3 of Ref.~\cite{regula2021operational} in the semidefinite programs, where $p_{\mathcal{Q}(A \rightarrow A)}$ is denoted as $\Vert \cdot \Vert_{\blacklozenge}$. 
    }

    \subsection{Logarithmic negativity}

    In Ref.~\cite{guo2023noise}, the relations of physical implementability to logarithmic negativity and purity are shown.
    In this section, we will generalize these results to the implementability function defined in this context.

    The logarithmic negativity of a quantum state $\rho$ of composited system $AB$ is defined as~\cite{PhysRevA.65.032314,PhysRevLett.95.090503}
    \begin{equation}
        E_N(\rho) = \log \Vert \rho^{T_B} \Vert_1,
    \end{equation}
    where $T_B$ is the partial transpose of subsystem $B$, and $\Vert \cdot \Vert_1$ denotes the trace norm.
    Here, we will show that the logarithmic negativity is related to the implementability function of composited system $AB$, when the free set $\mathcal{F}(AB)$ of states is all the physical states, namely the set $\mathcal{Q}(AB)$ of all positive semi-definite normalized Hermitian operators.
    \begin{lemma} \label{lem: lower-bound}
        Let the free set $\mathcal{F}(X) \subset \mathcal{Q}(X)$ of system $X$ is physical, the implementability function $p$ is lower bounded
        \begin{equation}
            p(\sigma) \geq \Vert \sigma \Vert_1
        \end{equation}
        for all $\sigma \in \mathcal{A}(X)$.
    \end{lemma}
    \begin{proof}
        Let the implementability function $p(\sigma)$ of the quasi-state $\sigma$ is attained as the decomposition
        \begin{equation}
            \sigma = n_1 \sigma_1 - n_2 \sigma_2,
        \end{equation}
        where $\sigma_1,\sigma_2 \in \mathcal{F}(X)$, $n_1,n_2 \geq 0$, and $p(\sigma) = n_1 + n_2$.
        Then, the trace norm of the quasi-state $\sigma$ is 
        \begin{align}
            \Vert \sigma \Vert_1 & = \Vert n_1 \sigma_1 - n_2 \sigma_2 \Vert_1 \leq n_1 \Vert \sigma_1 \Vert_1 + n_2 \Vert \sigma_2 \Vert_1 \nonumber\\
            & = n_1 + n_2 = p(\sigma) .
        \end{align}
        The first equality in the second line is because that $\sigma_1,\sigma_2 \in \mathcal{F}(X) \subset \mathcal{Q}(X)$ is positive semi-definite, and the trace norm of them are the trace $1$.
    \end{proof}
    \begin{proposition} \label{pro: trace_norm}
        The implementability function $p_{\mathcal{Q}}$ of a system $A$ with the free set $\mathcal{F}(A) = \mathcal{Q}(A)$ is the trace norm $\Vert \cdot \Vert_1$.
    \end{proposition}
    \begin{proof}
        The quasi state $\sigma \in \mathcal{A}(A)$, which is Hermitian, can be diagonalized as $\sigma = \sum_{i} p_i \ket{i} \bra{i}$, where $\sum_i p_i = 1$.
        The trace norm is $\Vert \sigma \Vert_1 = \sum_i |p_i|$.
        Since the free set $\mathcal{F}(A) = \mathcal{Q}(A)$, the orthogonal basis $\ket{i} \bra{i} \in \mathcal{Q}(A)$.
        By the definition of implementability function, 
        \begin{equation}
            p(\sigma) \leq \Vert \sigma \Vert_1,
        \end{equation} 
        and we prove what we want by combining with the Lemma~\ref{lem: lower-bound}.
    \end{proof}
    \begin{corollary}
        The logarithmic negativity of state $\rho$ is the logarithm of the implementability function $p(\rho^{T_B})$ of the partial transpose $\rho^{T_B}$ with respect to the free set $\mathcal{F}(A) = \mathcal{Q}(A)$
        \begin{equation}
            E_N(\rho) = \log p_{\mathcal{Q}}(\rho^{T_B}).
        \end{equation} 
    \end{corollary}\noindent
    This theorem shows the relation between logarithmic negativity and the implementability function.
    Moreover, since the partial transpose $T_B(\rho) = \rho^{T_B}$ is a linear involution on $\mathcal{B}(AB)$, $T_B^2 = \mathrm{id}$, with affine invariance, we have 
    \begin{equation}
        E_N(\rho) = \log p_{T_B(\mathcal{Q})}(\rho).
    \end{equation}
    It relates the negativity $N(\rho) = \frac{\Vert \rho^{T_B} \Vert_1 - 1}{2}$, which is an entanglement monotone without striking operational interpretation~\cite{PhysRevLett.95.090503}, to robustness $R_{T_B(\mathcal{Q})}(\rho)$ on the set of partial transpose of quantum states. 
    The above results inspire us to generalize the logarithmic negativity.
    \begin{definition} \label{def: negativity}
        The logarithmic negativity $E_{N \mathcal{F}}$ for $\rho \in \mathcal{B}(AB)$ with respect to a convex set $\mathcal{F}(AB)$ is defined as 
        \begin{equation}
            E_{N \mathcal{F}}(M) = \log p_{T_B(\mathcal{F})(AB)}(\rho) ,
        \end{equation}
        where $p_{T_B(\mathcal{F})(AB)}$ is the implementability function with respect to partial transpose $T_B(\mathcal{F})(AB)$ of the free set $\mathcal{F}$.
    \end{definition}
    \begin{proposition} \label{pro: negativity}
        Let $\rho = \mathcal{N}(\rho_0)$, where $\rho_0 \in \mathcal{A}(AB)$, $\mathcal{N} \in \mathcal{A}(AB \rightarrow AB)$
        \begin{align}
            E_{N \mathcal{F}(AB)}(\rho) & - E_{N \mathcal{F}(AB)}(\rho_0) \nonumber\\
            & \leq \log p_{T_B(\mathcal{F})(AB \rightarrow AB)}(\mathcal{N}). 
        \end{align}
        The inequality is tight when $\mathcal{F}(AB \rightarrow AB) = \mathcal{F}_{\max}(AB \rightarrow AB)$ is resource nongenerating set.
    \end{proposition}
    \begin{proof}
        \begin{align}
            E_{N \mathcal{F}(AB)}(\rho) & - E_{N \mathcal{F}(AB)}(\rho_0) = \log \frac{p_{\mathcal{F}(AB)}(\rho^{T_B})}{p_{\mathcal{F}(AB)}(\rho_0^{T_B})} \nonumber \\
            & = \log \frac{p_{\mathcal{F}(AB)}(T_B \circ \mathcal{N} \circ T_B(\rho_0^{T_B}))}{p_{\mathcal{F}(AB)}(\rho_0^{T_B})} \nonumber\\
            & \leq \log p_{\mathcal{F}(AB \rightarrow AB)}(T_B \circ \mathcal{N} \circ T_B) \nonumber \\
            & = \log p_{T_B(\mathcal{F})(AB \rightarrow AB)}(\mathcal{N}).
        \end{align}
        The last inequality is from the Corollary~\ref{cor: state-operation bound}, and the last equality is from the affine invariance, Corollary~\ref{cor: affine_invariant}.
        The tightness followed from Proposition~\ref{pro: diamond}.
    \end{proof}

    \begin{corollary}
        When the operation $\mathcal{N}$ commute with the partial transpose $T_B$, 
        \begin{equation}
            E_{N \mathcal{F}(AB)}(\rho) - E_{N \mathcal{F}(AB)}(\rho_0) \leq \log p_{\mathcal{F}(AB \rightarrow AB)}(\mathcal{N}).
        \end{equation}
        In particular, it holds for local operations
        \begin{equation}
            \mathcal{N} = \sum_i q_i \mathcal{N}_{iA} \otimes \mathcal{N}_{iB}
        \end{equation}
    \end{corollary}\noindent
    This generalizes the result in Ref.~\cite{guo2023noise}. 
    \begin{proof}
        The inequality follows directly from Proposition~\ref{pro: negativity}, if $\mathcal{N}$ commute with $T_B$.

        Since the density matrix $\rho_B$ of system $B$ is Hermitian, $\rho_B^{T_B} = \rho_B^*$.
        The linear operation $\mathcal{N}_B$ on system $B$ commutes with the transpose on system $B$
        \begin{align}
            & T_B \circ \mathcal{N}_B \circ T_B(\rho_B) = [\mathcal{N}_B(\rho_B^{T_B})]^{T_B} \nonumber \\
            &~~~~ = [\mathcal{N}_B(\rho_B^{*})]^{T_B} 
            = [\mathcal{N}_B(\rho_B)]^{\dagger} = \mathcal{N}_B(\rho_B).
        \end{align}
        The third equality from the linearity of the operation $\mathcal{N}$.
        For system $AB$, any state $\rho_{AB}$ can be decomposed as $\rho_{AB} = \sum_i x_i \rho_{iA}\otimes \rho_{iB}$.
        The operation on system $B$ completely commutes with the partial transpose $T_B$
        \begin{align}
            &\mathrm{id}_A \otimes (T_B \circ \mathcal{N}_B \circ T_B)(\rho_{AB}) \nonumber\\
            &~~~~ = \sum_i x_i \rho_{iA}\otimes T_B \circ \mathcal{N}_B \circ T_B(\rho_{iB}) \nonumber\\
            &~~~~ = \mathrm{id}_A \otimes \mathcal{N}_B(\rho_{AB}).
        \end{align}
        Thus, local operations $\mathcal{N}$ commute with partial transpose $T_B$
        \begin{align}
            & T_B \circ \mathcal{N} \circ T_B = \sum_i q_i \mathcal{N}_{iA} \otimes (T_B \circ \mathcal{N}_{iB} \circ T_B) \nonumber \\
            &~~~~ = \sum_i q_i \mathcal{N}_{iA} \circ [\mathrm{id}_A \otimes (T_B \circ \mathcal{N}_{iB} \circ T_B)] \nonumber \\
            &~~~~ = \sum_i q_i \mathcal{N}_{iA} \otimes \mathcal{N}_{iB} = \mathcal{N}.
        \end{align}
    \end{proof}

    \subsection{Purity}

        The purity is the square of the Frobenius (or Hilbert-Schmidt) norm $\Vert \cdot \Vert_2$, for convenience, we consider the ratio of Frobenius norms of initial quasi-state $\sigma$ and output state $\mathcal{N}(\sigma)$ in the following.
        Then, we view the space of state as a vector space, and the state matrices as vectors.
        In detail, we select an orthonormal basis $\ket{i}$, so the matrix basis is $\hat{M}_{i}^{\, j} \equiv \ket{i} \bra{j}$ with the multiplication rule $M_{i}^{\, j} M_{k}^{\, l} = \delta_{k}^{\, j} M_{i}^{\, l}$.
        With this basis, we expand the operations $\mathcal{N}$ as
        \begin{equation}
            \mathcal{N}(\sigma) = N^{i \, k}_{\, j \, l} \hat{M}_{i}^{\, j} \sigma \hat{M}_{k}^{\, l},
        \end{equation}
        and the density matrix as
        \begin{equation}
            \sigma = \sigma^{m}_{\, n} \hat{M}_{m}^{\, n}.
        \end{equation}
        By simple calculation, we get the ``matrix'' representation of operations on the operator space
        \begin{equation}
            \mathcal{N}(\sigma)^{i}_{\, j} = N^{i \, n}_{\, m \, j} \sigma^{m}_{\, n}.
        \end{equation}
        It is possible to rephrase this representation in the more familiar form of the vector~\cite{gilchrist2011vectorization}, which just raises the ``bra'' index of the density matrix, thus we are still working in the tensor here.

        For the ``matrix'' $N^{i \, n}_{\, m \, j}$, there exists a ``unitary'' transformation $U$, subordinate to the Hilbert-Schmidt inner product $(\hat{A}, \hat{B}) = \mathrm{Tr}(\hat{A}^{\dagger} \hat{B})$ of matrix space, transform the ``matrix'' $N^{i \, n}_{\, m \, j}$ into Jordan canonical form~\cite{vinberg2003course}, on some orthonormal matrix basis
        \begin{equation}
            N = U \bigoplus_{i} N_i U^{\dagger},
        \end{equation}
        where $N_i - n_i I$ is a nilpotent matrix
        \begin{equation}
            (N_i - n_i I)^{d_i} = 0,
        \end{equation}
        with $d_i$ the dimension of the matrix $N_i$.
        Therefore, we have 
        \begin{equation}
            N_i = n_i I + b_i P_i,
        \end{equation} 
        where 
        \begin{equation}
            P_i = \left(\begin{array}{cccc}
                0 & 1 & \\
                & 0 & \ddots & \\
                & & \ddots & 1   \\
                & & & 0
            \end{array}\right)
        \end{equation}
        is a nilpotent matrix of $d_i$ dimension, $b_i$ is a coefficient.
        With the Jordan canonical form, the state space is decomposed as the invariant subspace of the ``matrix'' $N^{i \, n}_{\, m \, j}$.
        In this decomposition, the state is decomposed as 
        \begin{equation}
            \sigma = \bigoplus_i \sigma_i, 
        \end{equation}
        so the ratio of Frobenius norm becomes
        \begin{equation}
            \frac{\Vert \mathcal{N}(\sigma) \Vert_2}{\Vert \sigma \Vert_2} = \frac{\sum_i \Vert \mathcal{N}_i(\sigma_i) \Vert_2 }{\sum_i \Vert \sigma_i \Vert_2} \leq \Vert \mathcal{N} \Vert_F,
        \end{equation}
        where the norm 
        \begin{equation}
            \Vert \mathcal{N} \Vert_F = \max_{i} \Vert \mathcal{N}_i \Vert_F = \max_{i} \max_{\sigma_i} \frac{\Vert \mathcal{N}_i(\sigma_i) \Vert_2}{\Vert \sigma_i \Vert_2},
        \end{equation}
        is induced by the Frobenius norm.
        Obviously, if $\mathcal{N}_i$ is diagonal, then $\Vert \mathcal{N}_i \Vert_F = |n_i|$, otherwise, $\Vert \mathcal{N}_i \Vert_F < |n_i| + |b_i|$.

        The norm $\Vert \mathcal{N} \Vert_F$ satisfies the triangle inequality, which is inherited from the Frobenius norm.
        Therefore, by selecting the optimal extreme-point decomposition with respect to the free set $\mathcal{F}$
        \begin{equation}
            \mathcal{N} = \sum_{l} n_l \mathcal{F}_l,
        \end{equation}
        we have
        \begin{equation} \label{eq: purity}
            \Vert \mathcal{N} \Vert_F \leq \sum_{l} |n_l| \Vert \mathcal{F}_l \Vert_F \leq p_{\mathcal{F}}(\mathcal{N}) \max_{l \in A^+ \cup A^-} \Vert \mathcal{F}_l \Vert_F.
        \end{equation}

        In conclusion, we have the following.
        \begin{proposition} \label{pro: purity}
            Let $\sigma \in \mathcal{A}(A)$, $\mathcal{N} \in \mathcal{A}(A \rightarrow A)$
            \begin{equation}
                \frac{\Vert \mathcal{N}(\sigma) \Vert_2}{\Vert \sigma \Vert_2} \leq p_{\mathcal{F}(A)}(\mathcal{N}) \max_{l \in A^+ \cup A^-} \Vert \mathcal{F}_l \Vert_F,
            \end{equation}
            In particular, if the operation $\mathcal{N}$ is unital, select $\sigma = \rho - I/D$, 
            \begin{equation}
                \log \frac{P(\mathcal{N}(\rho)) D - 1}{P(\rho) D - 1} \leq 2 \log p_{\mathcal{F}(A)}(\mathcal{N}) + 2 \log \max_{l \in A^+ \cup A^-} \Vert \mathcal{F}_l \Vert_F,
            \end{equation}
            Moreover, if $\mathcal{N}$ is unitary mixed whose unitaries are free, namely $\mathcal{F}_l$ are unitary, 
            \begin{equation}
                \log \frac{P(\mathcal{N}(\rho)) D - 1}{P(\rho) D - 1} \leq 2 \log p_{\mathcal{F}(A)}(\mathcal{N}),
            \end{equation}      
        \end{proposition}\noindent
        This generalizes the result in Ref.~\cite{guo2023noise}.
        \begin{proof}
            The first inequality is proved in Eq.~(\ref{eq: purity}).
            If the operation $\mathcal{N}$ is unital, select $\sigma = \rho - I/D$, then $\mathcal{N}(\sigma) = \mathcal{N}(\rho) - I/D$.
            For the purity $P(\rho) = \Vert \rho \Vert_2^2$, we have $P(\sigma) = P(\rho) - 1/D$.
            The second inequality is proved by substituting this in the first one.
            For unitary mixed channels, $\mathcal{F}_l$ are unitary.
            Since unitary channel is diagonal with eigenvalue $1$, $\max_{l \in A^+ \cup A^-} \Vert \mathcal{F}_l \Vert_F = 1$, which proves the third inequality.
        \end{proof}
        
%